\documentclass[longbibliography,twocolumn,prb,aps,superscriptaddress,showpacs,amsmath,amssymb,floatfix]{revtex4-2}
\usepackage{graphicx}
\usepackage{amssymb}
\usepackage{amsmath}
\usepackage{epsfig}
\usepackage{color}
\usepackage{floatrow}
\usepackage{flushend}

\usepackage{mathtools}
\usepackage{mathdots}
\usepackage{siunitx}
\usepackage[version=4]{mhchem}
\usepackage[colorlinks,linkcolor=blue,anchorcolor=blue,citecolor=blue,urlcolor=blue]{hyperref}
\usepackage{physics}
\usepackage{bm}
\graphicspath{{BLG_fig/}{figsgaoerb/}}

\begin{document}
	
	\title{Quasiparticle and Transport Properties of Disordered Bilayer Graphene}
	\author{Yanru Chen}
	\thanks{These authors contributed equally to this work}
	\affiliation{International Center for Quantum Design of Functional Materials (ICQD), Hefei National Research Center for Physical Sciences at the Microscale, University of Science and Technology of China, Hefei, Anhui 230026, China}
	\affiliation{Hefei National Laboratory, University of Science and Technology of China, Hefei, Anhui 230088, China}
	\author{Bo Fu}
	\thanks{These authors contributed equally to this work}
	\affiliation{Department of Physics, The University of Hong Kong, Pokfulam Road, Hong Kong, China}
	\affiliation{School of Sciences, Great Bay University, Dongguan, China}
	\author{Jinrong Xu}
	\affiliation{Key Laboratory of Advanced Electronic Materials and Devices, School of Mathematics and Physics, Anhui Jianzhu University, Hefei, Anhui 230601, China}
	\author{Qinwei Shi}
	\affiliation{International Center for Quantum Design of Functional Materials (ICQD), Hefei National Research Center for Physical Sciences at the Microscale, University of Science and Technology of China, Hefei, Anhui 230026, China}
	\author{Ping Cui}
	\thanks{Corresponding authors: 
		\\cuipg@ustc.edu.cn; zhangzy@ustc.edu.cn}
	\affiliation{International Center for Quantum Design of Functional Materials (ICQD), Hefei National Research Center for Physical Sciences at the Microscale, University of Science and Technology of China, Hefei, Anhui 230026, China}
	\affiliation{Hefei National Laboratory, University of Science and Technology of China, Hefei, Anhui 230088, China}
	\author{Zhenyu Zhang}
	\thanks{Corresponding authors: 
		\\cuipg@ustc.edu.cn; zhangzy@ustc.edu.cn}
	\affiliation{International Center for Quantum Design of Functional Materials (ICQD), Hefei National Research Center for Physical Sciences at the Microscale, University of Science and Technology of China, Hefei, Anhui 230026, China}
	\affiliation{Hefei National Laboratory, University of Science and Technology of China, Hefei, Anhui 230088, China}
	
	\begin{abstract}
		In recent experimental and theoretical studies of graphene, disorder scattering processes have been suggested to play an important role in its electronic and transport properties. In the preceding paper, it has been shown that the nonperturbative momentum-space Lanczos method is able to accurately describe all the multiple impurity scattering events and account for the quasiparticle and transport properties of disordered monolayer graphene. In the present study, we expand the range of applicability of this recursive method by numerically investigating the quasiparticle and transport properties of Bernal-stacked bilayer graphene in the presence of scalar Anderson disorder. The results are further compared with the findings of the same system using a self-consistent Born approximation, as well as the central findings in the preceding paper for monolayer graphene. It is found that in both systems, proper inclusions of all the scattering events are needed in order to reliably capture the role of disorder via multiple impurity scattering. In particular, the quasiparticle residue is shown to decrease sharply near the charge neutrality point, suggesting that the system is either a marginal Fermi liquid or a non-Fermi liquid. Furthermore, we reveal the dependences of the transport properties of disordered bilayer graphene on the carrier density and temperature, and explore the role of interlayer scattering at varying strengths. Our findings help to provide some new angles into the quasiparticle and transport properties of disordered bilayer graphene.
	\end{abstract}
	\maketitle
	
	\section{Introduction} 
	Graphene, a single carbon atomic layer, was experimentally realized in 2004 \cite{novoselov2004electric}. 
	It is the simplest two-dimensional Dirac material with its low energy excitation described by massless chiral Dirac Fermions.
	By exploiting the van der Waals interlayer coupling, various kinds of bilayer graphene (BLG) have been realized, such as Bernal-stacked bilayer graphene with a parabolic dispersion at low energies \cite{mccann2013electronic}, and magic-angle twisted bilayer graphene with strongly coupled flat bands \cite{cao2018unconventional,cao2018correlated}.
	As presented on the preceding paper, extensive experimental studies on the electronic transport properties of monolayer graphene (MLG) \cite{mayorov2012close,novoselov2005twodimensional,bolotin2008temperature,morozov2008giant,du2008approaching,tan2007measurement,dean2010boron,zomer2011transfer} and Bernal-stacked bilayer graphene  \cite{novoselov2006unconventional,morozov2008giant,dean2010boron,zomer2011transfer,katoch2018transport,feldman2009brokensymmetry} have shown that disorder scattering plays an important role in these systems, and the different types of disorder may dominate under different physical conditions. 
	For example, for graphene on the \ce{SiO_2} substrate, electron-hole puddles have been suggested as the dominant source of remnant disorder in the form of long-range charged impurities \cite{martin2008observation,hwang2007carrier}. For suspended graphene and graphene encapsulated in hexagonal boron nitride layers \cite{dean2010boron}, the combination of long-range and short-range impurities has been invoked to explain their sublinear behavior of conductivity \cite{dassarma2010theory,dassarma2011conductivity}. But shortage of experimental evidence for the existence of short-range impurities in monolayer graphene still questions the role of scalar Anderson impurities. On the other hand, for bilayer graphene, native point-like defects have recently been detected by using scanning tunneling microscopy, calling for closer attention to the potential  importance of scalar short-range disorder in few-layer graphene systems \cite{Joucken_2021_sublattice,joucken2021direct}. 
	
	On the theory side, diagrammatic methods such as the self-consistent Born approximation are commonly used to study disorder physics \cite{groth2009theory,bruus2004manybody}. The diagrammatic approximation is useful when a particular set of diagrams plays a major role and is intuitive in understanding the intrinsic physical processes. However, it is often challenging to identify and calculate the dominant types of diagrams in a given complex system. In particular, coherent multiple scattering of electrons in disordered materials can cause a large number of physical phenomena such as Anderson localization, weak (anti-)localization, and universal conductance fluctuation \cite{pixley2015andersona,gorbachev2007weak,ilic2019weak}, while proper descriptions of such phenomena are extremely demanding within standard diagrammatic approaches. Moreover, when referring to few-layer graphene, the commonly used self-consistent Born approximation neglects  many multi-impurity scattering events, yet such events may become important in Dirac materials even for weak disorder \cite{vanrossum1999multiple,zhu2010evaluation}. Therefore, the non-perturbative Lanczos method \cite{lanczos1950iteration,senechal2010introduction} is needed in studying the disorder effects in these systems. 
	This method has been widely used to study many-body effects \cite{jaklic1994lanczos,becca2000groundstate,tanaka2019metalinsulator} and  disorder physics \cite{zhu2010evaluation,zhu2012vacancyinduced,Fu_2017_Accurate}. More importantly, the multiple impurity scattering processes involving different impurity centers can be treated exactly using this method \cite{zhu2010evaluation,ning2020multiscattering,chen2020spinorbit}, a capability that is particularly valuable for investigating disordered bilayer graphene that contains both intralayer and interlayer scattering events.

	In this paper, we study the quasiparticle and transport properties of bilayer graphene with short-range Anderson disorder using the Lanczos method. The quasiparticle properties are studied in both the strong ($ E_f\tau/\hbar \leq 1 $) and weak scattering limits ($ E_f\tau/\hbar \gg 1 $), with $ E_f $ the Fermi energy and $ \tau $ the quasiparticle lifetime. In particular, the quasiparticle residue is shown to decrease sharply near the charge neutrality point, suggesting the modification of multiple scattering. Furthermore, we find that the conductivity increases with the carrier density and saturates at high carrier densities, and the interlayer scattering events will reduce the longitudinal conductivity. We also obtain the characteristic dependence of the conductivity on the temperature in the low carrier density limit (namely, around the charge neutrality point). The results are further compared with the findings of the same system using the self-consistent Born approximation, demonstrating the pronounced differences between the two approaches, as well as the central findings in the preceding paper for monolayer graphene, highlighting the interlayer scattering effects. 
	
	This paper is organized as follows. The tight-binding model and methodologies are introduced in Sec. \ref{sec.model}. The numerical results for quasiparticle properties calculated by the Lanczos method and SCBA are presented in Sec. \ref{sec.result}. The transport properties are given in Sec. \ref{sec.transport}. Finally, in Sec. \ref{sec.discussion}, a brief conclusion is given.
	
	\section{Model and methods}
	\label{sec.model}

\begin{figure}[t!]
		\centering
		\includegraphics[width=1\linewidth]{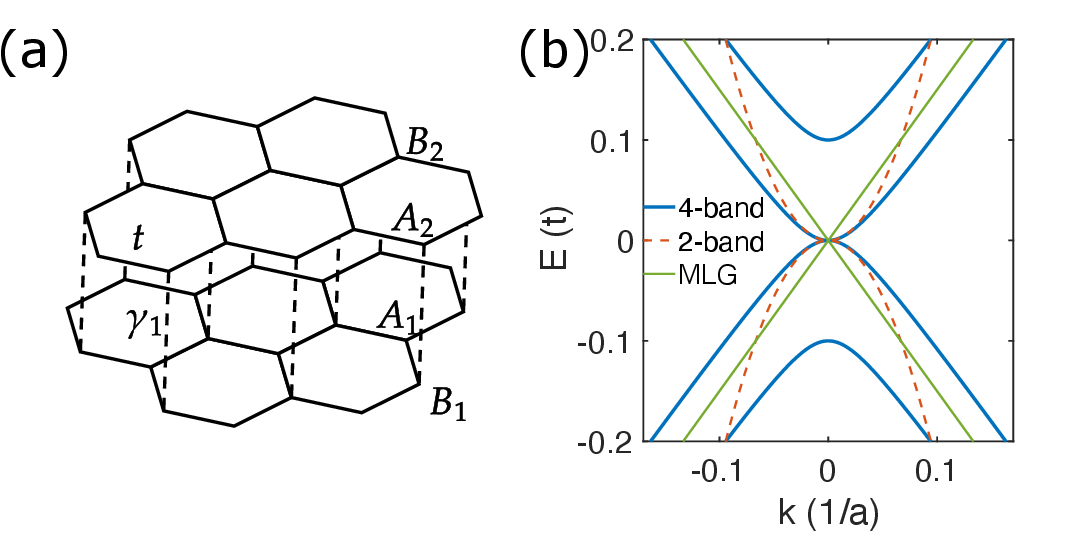}
		\caption{(a) Schematic of a Bernal-stacked bilayer graphene structure, with $ A_1$ and $B_1 $ ($ A_2$ and $B_2 $) indicating the sublattices on the bottom (top) layer. The solid (dashed) lines represent the nearest intralayer (interlayer) hopping $ t $ ($ \gamma_1 $). (b) Band structures without disorder around the $ \text{K}_\pm $ point. The blue solid and orange dashed lines represent the spectrum of BLG within the four-band and two-band models, respectively. The solid green lines are the spectrum of monolayer graphene.}
		\label{fig.structure}
	\end{figure}

	We start with the tight-binding model for clean Bernal-stacked (AB-stacked) BLG \cite{mccann2013electronic} (see Fig.~\ref{fig.structure}(a))
	\begin{align}
		H_0&=-t \sum_{b} \sum_{\langle i,j\rangle} c_{b,i}^\dag c_{b,j} +\gamma_1 \sum_i c_{\text{A}_2,i}^\dag c_{\text{B}_1,i} ,
	\end{align}
	where $ c^\dag $ ($ c $) is the creation (annihilation) operator, $b = \text{A}_1, \text{B}_1,\text{A}_2, \text{B}_2$ refer to sublattices ($``1"$ and $``2" $ refer to the bottom and top layers, respectively), $ i$ and $j $ represent the coordinates of unit cells. The nearest intralayer and interlayer hoppings are denoted by $ t $ and $ \gamma_1 $, respectively. Other additional interlayer hopping terms are neglected since they are much smaller than $ \gamma_1 $. When $ \gamma_1=0 $, the system becomes two decoupled graphene monolayers. To consider a moderate and reasonable interlayer hopping, we choose $ \gamma_1=0.1t $ in our calculations \cite{mccann2013electronic}. For convenience, we set all the energies (bands, hopping strength, etc) in the unit of $ t=\SI{2.7}{eV}  $ and the length in the unit of lattice constant (carbon-carbon distance) $ a=\SI{0.142}{nm} $ in the whole paper.

	The corresponding Hamiltonian $ H_0 $ can be solved in the momentum space near the valley $\mathbf{K}_\xi=(\xi \frac{4\pi}{3\sqrt{3}a},0) $, with valley index $ \xi=\pm $ denoting the two nonequivalent valleys, which reads as
	\begin{align}
		H_{K_{\xi}}= \begin{pmatrix}
			0 & \hbar v_f k_-& 0 & 0 \\
			\hbar v_f k_+ & 0 & \gamma_1 & 0\\
			0& \gamma_1 & 0 & \hbar v_f k_-\\
			0 & 0 & \hbar v_f k_+& 0
		\end{pmatrix},
	\end{align}
	where $ k_\pm= \xi k_x\pm ik_y $, and $ v_f=3at/2\hbar $ is the Fermi velocity of MLG. The basis is chosen as $ \psi=(c_{\text{A}_1,\bm{k}}, c_{\text{B}_1,\bm{k}}, c_{\text{A}_2,\bm{k}}, c_{\text{B}_2,\bm{k}})^T $, with $ c_{b,\bm{k}} $ ($ c_{b,\bm{k}}^\dag $) being the annihilation (creation) operator in the momentum space. The corresponding spectrum is then obtained as $ \varepsilon(\bm{k}) =\pm \left[ \sqrt{(\hbar v_f k)^2 + (\gamma_1/2)^2} \pm (\gamma_1/2) \right] $ with $ k=|\bm{k}| $, exhibiting the four bands as plotted in Fig.~\ref{fig.structure}(b) (the blue solid lines). Due to the interlayer coupling, the spectrums are parabolic in the vicinity of $ E=0 $ and recover to linear dispersion at larger energies, which is different from the massless Dirac cone of MLG (the green solid lines in Fig.~\ref{fig.structure}(b)).
	
	In the low energy regime $ |E|\ll \gamma_1/4 $ 
	, by projecting onto the lowest energy orbits, the four-band model mentioned above can be reduced to an effective two-band model with the basis $ \psi=(c_{\text{A}_1,\bm{k}}, c_{\text{B}_2,\bm{k}})^T $, and accordingly, the Hamiltonian is written as
	\begin{align}
		H_0^{\text{eff}}= -\frac{1}{2m} \begin{pmatrix}
			0 & k_-^2 \\
			k_+^2 & 0
		\end{pmatrix},
	\end{align}
	where the effective mass $ m=\gamma_1/(2 \hbar^2 v_f^2) $. The spectrum of the two-band model is $ \varepsilon_{\pm} = \pm k^2/(2m) $, shown as the orange dashed lines in the Fig.~\ref{fig.structure}(b).  Obviously, the two-band and four-band models are consistent perfectly within the small energy regime near the charge neutrality point (CNP, $ E=0 $).

	To explore the disorder effect, the short-range Anderson type nonmagnetic disorder is introduced by random on-site delta potential
	\begin{align}
		V(\bm{r})=\sum_i u_i \delta(\bm{r}- \bm{R}_i),
	\end{align}
	where $ u_i $ measures the random potential at position $ \bm{R}_i $  distributed uniformly and independently within the interval $ [-W/2, W/2] $.  And the correlation between impurities is $ \langle V(\bm{r}) V(\bm{r}^\prime)\rangle =n_{\text{imp}} u^2= n_{\text{imp}} A_c\frac{ W^2}{12} \delta(\bm{r}-\bm{r}^{\prime}) $. Here, $ n_{\text{imp}}=N_{\text{imp}}/N $ is the concentration of impurity, $ A_c=\frac{3\sqrt{3}}{2}a^2 $ is the area of the unit cell. A dimensionless parameter $ u_0=\frac{n_{\text{imp}} A_c W^2}{12(\hbar v_f)^2\pi} $ is defined to character the disorder strength.  Assume that the disorder is uncorrelated between sublattices due to its short-range nature. The disorder strength can be modified by adjusting $ W $ for ﬁxed $ n_{\text{imp}} =1 $. The range of disorder strengths considered in this paper is $ 0.02 \leqslant u_0 \leqslant 0.17 $ (which corresponds to $0.81t \leqslant W \leqslant 2.36t $).

	Here, we use the Lanczos recursive method in both the momentum space and real space to numerically compute the quasiparticle properties of BLG, including the exact ensemble-averaged retarded Green's function $ G^R $, the self-energy $ \Sigma $, and density of states (DOS). We also compare the self-energies obtained by the self-consistent Born approximation and Lanczos recursive method.  In order to avoid the finite-size effects, we choose a large sample containing millions of atoms ($ N=4\times3600^2 $). Moreover, a small artificial cutoff $ \eta=10^{-3} $ is used to simulate the infinitesimal imaginary energy in our calculations. At last, the periodic boundary condition is satisfied.

	\section{Quasiparticle Properties}
	\label{sec.result}
	
	\subsection{Self-energy}
	The renormalization for the single electron due to impurity scattering is encoded in the self-energy. Based on the Dyson equation $ G(\bm{k},E)= G_0(\bm{k},E) + G_0(\bm{k},E) \Sigma(\bm{k},E) G(\bm{k},E)$  \cite{mahan2000manyparticle}, the self-energy is defined as
	\begin{align}
		\Sigma(\bm{k}, E)= G^{-1}_{0} (\bm{k}, E) -G^{-1} (\bm{k}, E).
	\end{align}
	Here, $ G_0 $ is the retarded Green's function of the bare Hamiltonian without disorder. The ensemble-averaged Green's function is given by $ G=\langle E-H_0-V+i\eta \rangle ^{-1} $, 
	where $ \langle \cdots \rangle $ indicates the average expected value over the random disorder configurations.
	
	\begin{figure}[t!]
		\centering
		\includegraphics[width=1\linewidth]{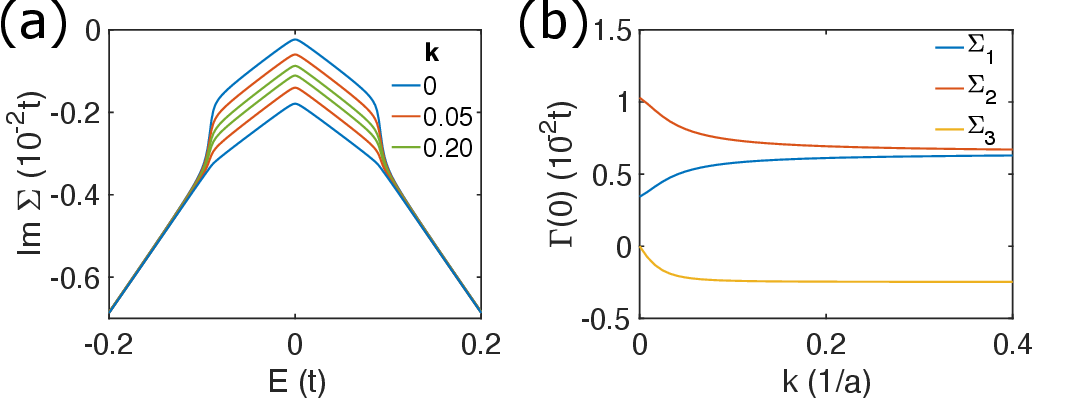}
		\caption{(a) $ \bm{k} $ and $ E $ dependences of the imaginary part of the self-energy $ \Im \Sigma(\bm{k},E) $ in the eigenstate representation. (b) $ \bm{k} $ dependence of $ \Gamma(0)=-\Im \Sigma(E=0) $. Here the disorder strength $ u_0=0.02 $. }
		\label{fig.k_Sigma}
	\end{figure}

	\subsubsection{Eigenstate Representation}
	The electronic properties are frequently addressed in the eigenstate representation so that the effects on the energy bands can be seen directly.  In the eigenstate representation, the self-energy is not diagonal and is dependent on both the energy $ E $ and wave vector $ \bm{k} $, and can be written as
	\begin{align}
		\Sigma(\bm{k},E)=\begin{pmatrix}
			\Sigma_1 & 0 & \Sigma_3 & 0 \\
			0 & \Sigma_2 & 0 & \Sigma_3 \\
			\Sigma_3 & 0 & \Sigma_2 & 0 \\
			0 & \Sigma_3 & 0 & \Sigma_1
		\end{pmatrix},
	\end{align}
	where the $ i $-th diagonal element can be considered as the self-energy of the $ i $-th energy band, and the detailed analysis is performed in the Appendix \ref{sec.SCBA}. The imaginary parts of the self-energy are shown in Fig.~\ref{fig.k_Sigma}. At high energies, the self-energy is wave vector $ \bm{k} $ independent. At low energies, the value of the self-energy element $ \Sigma_1 $ ($ \Sigma_2 $) increases (decreases) with the increase of $ k $, leading to the overlap of the two self-energy elements. Within a small energy range $ | E | \ll\gamma_1/4 $ ($k\ll \gamma_1/2v_f$), we can assume that the self-energy is momentum independent and take a $ \bm{k}=0 $ approximation. The $ \bm{k} $ independence assumption is also used with the effective homogeneous medium in perturbation calculations  \cite{altland2010condensed}. When $ \gamma_1=0 $, the result is consistent well with the case of MLG \cite{zhu2010evaluation}. It is obvious that the self-energy of MLG is only dependent on the energy $ E $, and not on the wave vector $ \bm{k} $.
	
	\begin{figure}[t!]
		\centering
		\includegraphics[width=0.95\linewidth]{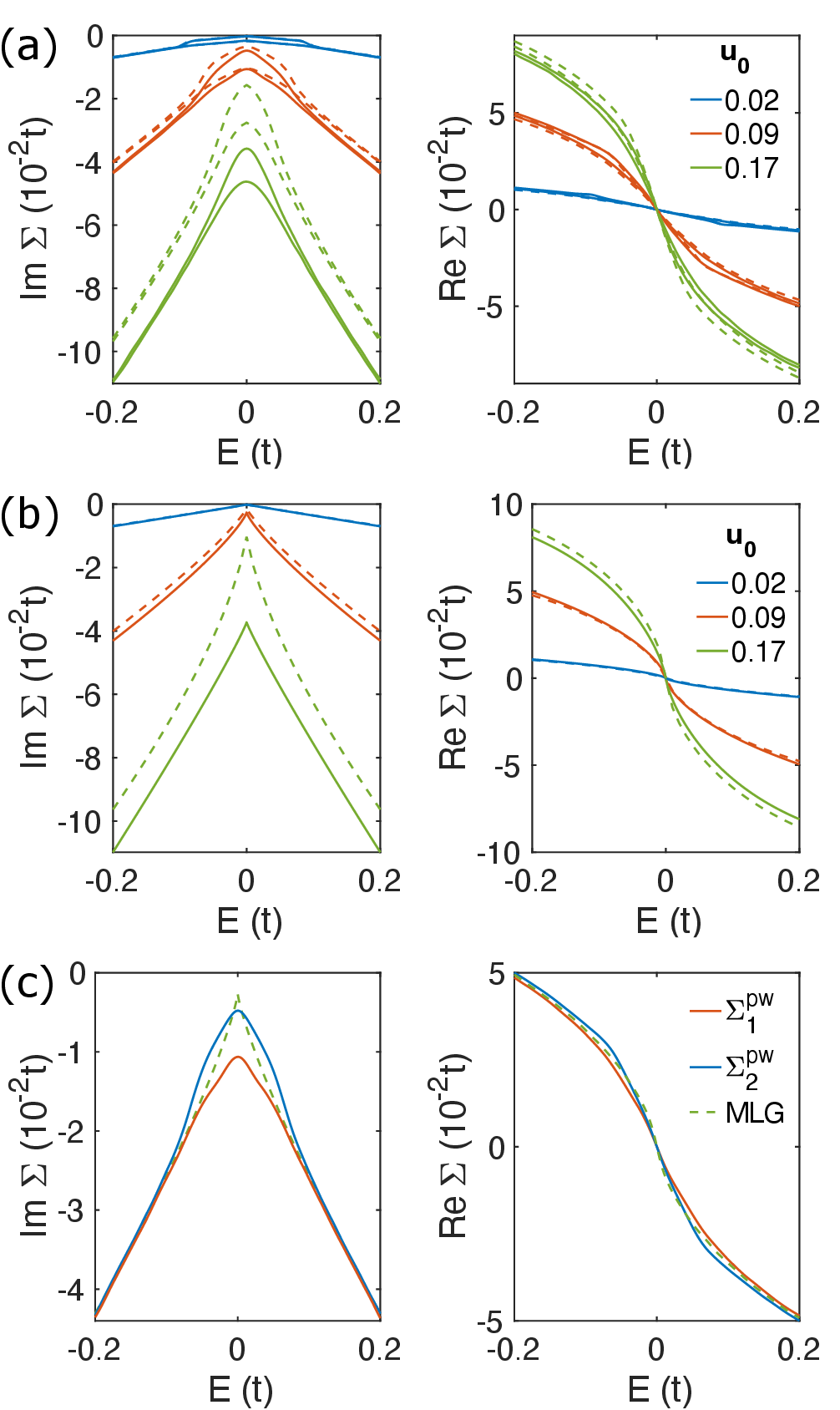}
		\caption{Energy dependences of the imaginary (left panel) and real  (right panel) parts of the self-energy elements for disordered (a) bilayer and (b) monolayer graphene in the plane wave representation. The solid (dashed) lines are the calculated results by the Lanczos method (self-consistent Born approximation) with different disorder strengths $ u_0=0.02,0.09, 0.17 $. (c) Comparison of the self-energies for the two systems at $ u_0=0.09 $. The red (blue) solid lines are the results of $ \Sigma^{\text{pw}}_1 $ ($ \Sigma^{\text{pw}}_2 $) for BLG, while the green dashed lines are the results for MLG.}
		\label{fig.Sigma}
	\end{figure}
	
	\subsubsection{Plane Wave Representation}
	 Another commonly used representation is the plane wave representation. In the plane wave representation, the self-energy is diagonal and momentum independent, and adopts the form
	 \begin{align}
	 	\Sigma^{\text{pw}} (E)=\begin{pmatrix}
	 		\Sigma_1^{\text{pw}} & 0 & 0 & 0 \\
	 		0 & \Sigma_2^{\text{pw}} & 0 & 0 \\
	 		0 & 0 & \Sigma_2^{\text{pw}} & 0 \\
	 		0 & 0 & 0 & \Sigma_1^{\text{pw}}
	 	\end{pmatrix}.
	 \end{align}
 	 For convenience, the subsequent discussion defaults to the plane wave representation. And the representation transformation between the plane wave and eigenstate representations is shown in the Appendix \ref{sec.transform}. 

\begin{figure}[t!]
		\centering
		\includegraphics[width=1\linewidth]{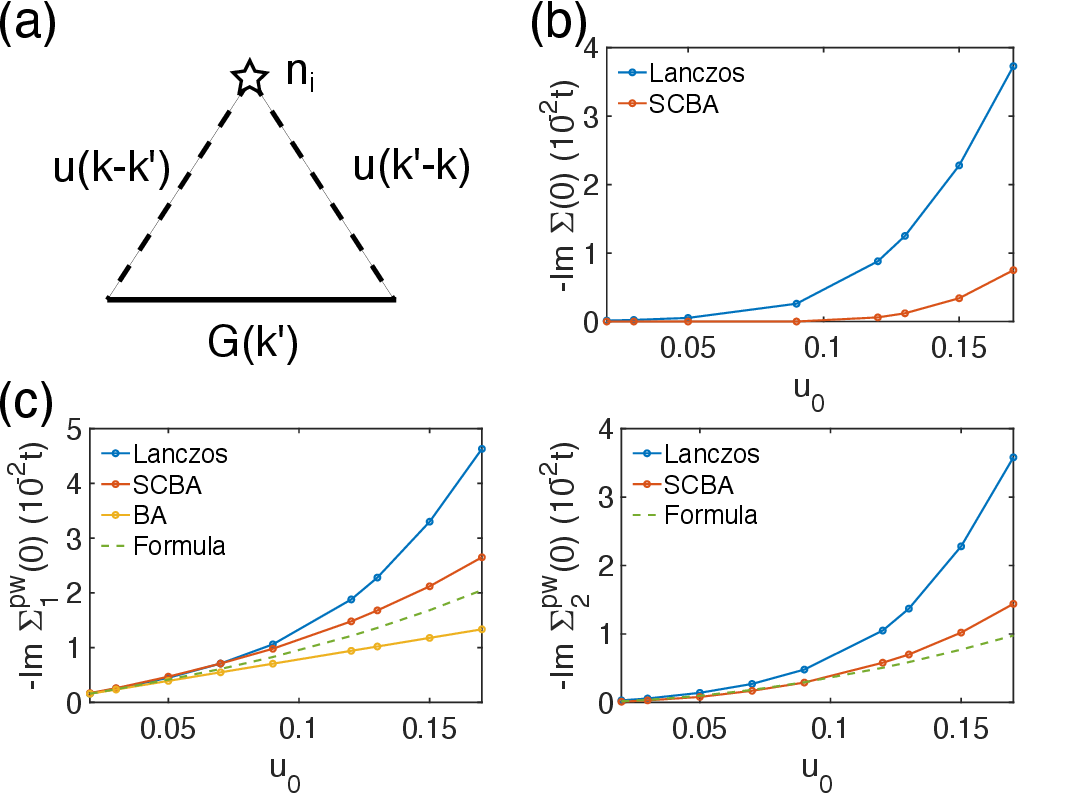}
		\caption{(a) Feynman diagram of the disorder averaged self-energy in the self-consistent Born approximation. The solid line represents Green's function, the star denotes impurity, and the dashed lines represent electron-impurity interaction. (b) Disorder strength dependence of the imaginary part of the self-energy of monolayer graphene at $ E=0 $. (c) Disorder strength dependence of $ \Im \Sigma_1^{\text{pw}} (0)$ (left panel) and $ \Im \Sigma_2^{\text{pw}} (0)$(right panel) of bilayer graphene. The blue (red) solid lines are the results calculated by the Lanczos method (SCBA). The yellow line is the result calculated by the Born approximation (BA) within the two-band model, which is the linear term of Eq. (\ref{equ.ImS01}). The green dashed lines are the analytic results of Eqs. (\ref{equ.ImS01}) and (\ref{equ.ImS02}).}
		\label{fig.ImS0}
	\end{figure}

 The energy-dependence imaginary part and real part of the self-energy for BLG with different disorder strengths $ u_0 $ are presented in Fig.~\ref{fig.Sigma}(a). 
	Since the imaginary part and real part can be transformed to each other via the Kramers-Kronig relation \cite{ref1}, we can only pay attention to the imaginary part $ \Im\Sigma $. 
	As shown in Fig.~\ref{fig.Sigma}(a), the amplitudes of $ \Im \Sigma_1 $ and $ \Im \Sigma_2 $ exhibit the same variation trend, both increasing with the dimensionless disorder strength $ u_0 $. Even though both the Lanczos method and SCBA (the diagram of the SCBA is shown in Fig.~\ref{fig.ImS0}(a)) can capture this feature, the discrepancy between the calculated results of the two methods is obvious, especially near the CNP. These two methods fit better when the impurity strength is weak, because the SCBA does not encompass all of the impurity effects well when the impurity strength increases. The amplitudes of the imaginary part $ \Im\Sigma $ within the SCBA are much smaller than the accurate results from the numerical simulations for strong disorder strength, indicating that multiple scattering plays an important role in BLG. The inaccuracy of the SCBA can be attributed to the mixture of Bloch states and the interference correlations from multiple scattering \cite{zhu2010evaluation,aleiner2006effect}.

	As a comparison, the self-energy for MLG is shown in Fig.~\ref{fig.Sigma}(b), which is momentum independent. The imaginary part of the self-energy follows a power law formula, and its amplitude also increases with the disorder strength $ u_0 $. The imaginary and real parts of the self-energies for two systems at $ u_0 = 0.09 $ are also contrasted in Fig.~ \ref{fig.Sigma}(c). It can be seen that the imaginary parts of the self-energies significantly differ near the CNP but are the same at higher energies, consistent with the band dispersion relations.

	To further investigate the effect of the disorder strength on the self-energy, we focus on the energy $ E=0 $ and obtain the analytic expression of the self-energy by the SCBA as
	\begin{align}
		\label{equ.ImS01}
		\Sigma^{\text{pw}}_1(E=0) &=-i\left( \frac{\pi\gamma_1}{4}u_0 +\frac{\pi \gamma_1 u_0^3}{16} \ln^2 \frac{4E_c^2}{\pi\gamma^2_1 u_0} \right),
		\\ \Sigma^{\text{pw}}_2 (E=0) &=-i \frac{\pi\gamma_1 u_0^2}{8} \ln \frac{4E_c^2}{\pi \gamma_1^2 u_0},
		\label{equ.ImS02}
	\end{align}
	here $ E_c=2.7t $ is the high energy cutoff. The first term of Eq.~(\ref{equ.ImS01}) is a linear relationship with the disorder strength $ u_0 $. 
	It is noted that we only get this linear term if we use the two-band model and ignore the contribution from more distant energy levels, which is also the Born approximation (BA). The second term of Eq.~(\ref{equ.ImS01}) and Eq.~(\ref{equ.ImS02}) relate to the coherence between the energy bands. As shown in Fig.~\ref{fig.ImS0}(c), the difference between the Lanczos and SCBA methods is insignificant when the disorder strength $ u_0 $ is weak. When $ u_0 $ increases, such difference increases, suggesting that the multiple scattering becomes important. For MLG, the self-energy obtained by the SCBA is written as $ \Sigma(E=0)= -iE_c \exp (-1/u_0) $. It is clear that the SCBA does not capture all the disorder effects very well, as shown in Fig.~\ref{fig.ImS0}(b). The multiple scattering effect is even more important in MLG than in BLG.
	Since the SCBA sums all the non-crossing diagrams, we wonder if the difference between the two methods is due to the cross terms. According to some previous studies, the SCBA is unreliable in semi-metals, where the condition $ k_Fl\leq1 $ ($ k_F $ the Fermi momentum and $ l $ the mean free path) is not satisfied \cite{aleiner2006effect,ostrovsky2006electron,sbierski2014quantum,nersesyan1995disorder}. In those systems, the second-order cross term has the same order magnitude as the SCBA results.
	However, for the short-range Anderson impurity used in this paper, the contribution of the second-order cross term (the lowest-order multi-scattering event) is zero, so the discrepancy should be attributed to the contribution of other higher-order Feynman diagrams.

	 \subsection{Density of States}
\begin{figure}[t!]
		\centering
		\includegraphics[width=1\linewidth]{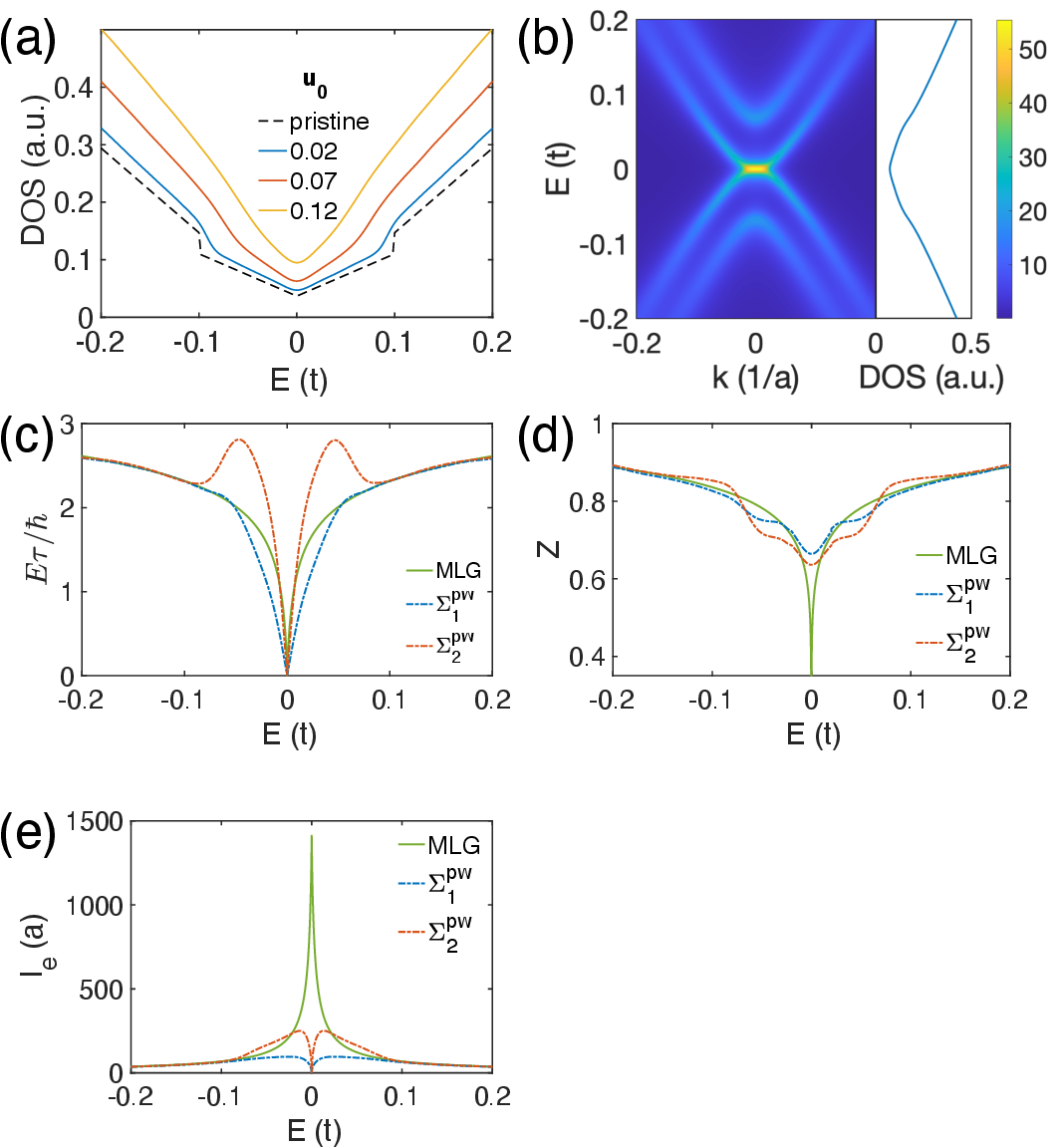}
		\caption{(a) Density of states (DOS) as a function of energy $ E $ with different disorder strengths $ u_0 $. The dashed (solid) lines are the results for the clean (disordered) bilayer graphene. (b) Spectral function $ A(\bm{k},E)$ and the corresponding DOS. (c) Dimensionless parameter $ E\tau/\hbar $ and (d) quasiparticle residue $ Z $ as functions of energy $ E $ at $ u_0=0.09 $. (e) Elastic mean free path $l_e$ as a function of energy $E$ at $u_0=0.02$. In (c)-(e), the solid and dashed lines are for MLG and BLG, respectively.}
		\label{fig.quasi}
	\end{figure}

	 A common and effective way to observe the disorder effect is to examine the changes of density of states (DOS) in the presence of the disorder. First, the DOS of clean BLG is given by
	\begin{align}
		D(E)= \frac{g_v g_s}{2\pi (\hbar v_f)^2}\left[|E| +\frac{\gamma_1}{2} +\Theta(|E| -\gamma_1) (|E|- \frac{\gamma_1}{2}) \right],
	\end{align}
	where $ \Theta(x) $ is the step function, $ g_s=2 $ and $ g_v=2 $ account for the spin degeneracy and valley freedom, respectively. Unlike MLG, BLG has a finite DOS at the CNP in addition to the linear dependence on the energy. 
	With the disorder, the DOS of a system can be provided by the imaginary part of the disorder averaged Green's function.
	In Fig.~\ref{fig.quasi}(a), we show the results of the DOS per unit cell obtained by the Lanczos method in the real space. It is found that the disorder significantly modifies the DOS and a relatively strong disorder can erase the step of the DOS. The imaginary part of the self-energy has similar behaviors with DOS, while the real part of the self-energy represents the renormalization of the energy bands. Combined with the changes of the real part of the self-energy and DOS, we can imagine that the energy levels are pushed toward zero energy by the disorder potential, and the DOS increases accordingly \cite{mccann2013electronic}. The increase in the DOS near the CNP is also evidence of multiple scattering.
	The short-range disorder does not qualitatively change the band structure. We plot the experimentally testable single-particle spectral function $ A(\bm{k},E)=-\Im G(\bm{k},E) /\pi $ as a function of energy $ E $ and momentum $ \bm{k} $ in Fig.~\ref{fig.quasi}(b). Naturally, the band structure is basically maintained, and more information is available in the Appendix. \ref{app.spectral}.

	\subsection{Quasiparticle residue}
	To measure the disorder effect on quasiparticle behaviors, we study the quasiparticle residue 
	\begin{align}
		Z=\left[1- \frac{\partial \Re \Sigma(\bm{k},E)}{\partial E}\right]^{-1} \bigg|_{E=\tilde{E}_{\bm{k}}},
	\end{align}   
	where $ \tilde{E}_{\bm{k}} $ is the energy of quasiparticle that is the root of the equation $ E-\varepsilon_{\bm{k}}- \Re \Sigma(\bm{k},E)=0 $. The quasiparticle residue is a crucial quantity to judge whether the system can be described by normal Fermi liquid (FL) theory \cite{zhao2016interplay}. If $ Z\approx 1 $, the system is close to a clean system, and can be well described by the normal FL. If $ Z $ is significantly smaller than $ 1 $ or even vanishes, it means that the system deviates from the original structure and the hybridization with other states is strong, indicating that the perturbation calculations are invalid and the system is a marginal FL or a non-FL \cite{zhu2010evaluation,Fu_2017_Accurate,zhao2016interplay}.  In addition, the elastic mean free time is defined as $ \tau=\hbar/ [-2Z\Im \Sigma(E)] $, the group velocity is $v_g= \partial E_k/\hbar \partial k= Z v$ with velocity $v=\partial \varepsilon_k/ \hbar \partial k$, and the elastic mean free path is  $l_e=v_g \tau= \hbar v/[-2 \Im \Sigma(E)]$. 
	As shown in Figs.~\ref{fig.quasi}(c) and ~\ref{fig.quasi}(d), in the weak scattering limit ($ E_f \tau/\hbar \gg 1 $), the behaviors of the elastic mean free time and quasiparticle residue for MLG and BLG are similar, with $ Z $ close to $ 0.9 $, and $Z$ decreases slowly as the energy $|E|$ decreases for each system. In the strong scattering limit ($ E_f \tau/\hbar \leq 1 $), $ Z $ further decreases,  especially for MLG where $Z$ drops rapidly at the CNP. Such unusual feature suggests that the multiple scattering effect remarkably changes the quasiparticle properties near the CNP. Furthermore, the behaviors of elastic mean free paths for two systems are contrasted in Fig.~\ref{fig.quasi}.(e). In the weak scattering region, the mean free paths for two system are nearly the same. However, in the strong scattering region, the mean free path for MLG becomes significantly longer as the $|E|$ decreases, even up to $10^3$ orders of magnitude, but the mean free path for BLG first increases and then decreases to zero. In the latter case, since the difference of the imaginary parts of the self-energies between the two systems is not significant, the difference of the elastic mean free path mainly comes from the electrons velocity. For MLG, the velocity of the electrons is $v_f$, while the velocity for BLG decreases as the energy approaches zero.

	\section{Transport Properties} 
	\label{sec.transport}
	
	\subsection{Longitudinal Conductivity}
	\begin{figure}[t!]
		\centering
		\includegraphics[width=1 \linewidth]{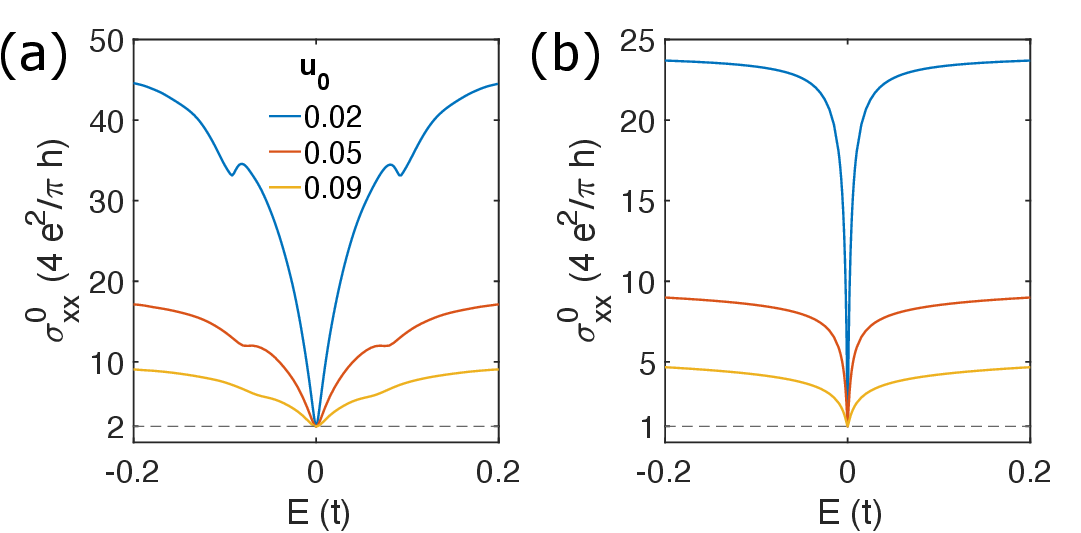}
		\caption{Conductivity at zero temperature ($ \sigma^0_{xx} $) for (a) BLG and (b) MLG with different disorder strengths $ u_0=0.02,0.05,0.09 $.}
		\label{fig.condE}
	\end{figure}

	To further study the disorder effect on transport properties of the system, we calculate the longitudinal conductivity  based on the Kubo-Greenwood formula \cite{akkermans2007mesoscopic}
	\begin{align}
		&\sigma_{xx} (E_f,T)= \int dE\ \left(-\frac{\partial f(E,E_f)}{\partial E}\right) \ \sigma^0_{xx}(E),
	\end{align}
	where $ f(E, E_f)=1/[e^{(E-E_f)/{k_B T}}+1] $ is the Fermi-Dirac distribution with $ E_f $ the Fermi energy, $ k_B $ the Boltzmann constant, and $ T $ the temperature, and $ \sigma^0_{xx} (E)$ is the zero temperature conductivity. At zero temperature the $ -\frac{\partial f(E,E_f)}{\partial E} $ can be replaced by delta function $ \delta(E-E_f) $. The zero-temperature conductivity is given by
	\begin{align}
		\sigma^0_{xx} (E)=g_s g_\nu \frac{e^2\pi\hbar}{L^2} \int \frac{d^2 \bm{k}}{(2\pi)^2} \Tr \left[v_{x} A(\bm{k},E) v_{x} A(\bm{k},E)\right],
	\end{align}
	with the velocity operator $ v_x= \frac{1}{\hbar} \frac{\partial H}{\partial k_x} $. Figure ~\ref{fig.condE}(a)(b) show the energy dependence of zero-temperature conductivities for BLG and MLG, respectively. Similar to that of a normal metal, the increase of the disorder strength leads to a decrease in the conductivity.
	The conductivity of either system increases extremely rapidly with increasing $ |E| $ and then saturates. One difference is that BLG has a kink around $ |E|=\gamma_1 $. The reason for the kink-like structure in the conductivity is the sudden appearance of interband scattering and additional carriers in the excited conduction band \cite{ando2011bilayer,mccann2013electronic}. Due to the broadening of the spectral function, the kink is gradually smeared when the disorder strength increases. Here, there also exists a minimum conductivity $ \sigma_{\text{min}} $ at the CNP for either system.

	In order to more intuitively study the disorder effects in the vicinity of $ E=0 $, we adopt the effective two-band model in the low energy regime $ |E|\ll \gamma_1/4 $. The zero-temperature conductivity near the CNP is expressed as
	\begin{align}
		\sigma^0_{xx} =\frac{4e^2}{\pi h} \left[1+(\frac{\alpha}{\Gamma} +\frac{\Gamma}{\alpha}) \arctan \frac{\alpha}{\Gamma}\right],
	\end{align}
	where $ \alpha=E-\Re \Sigma $ and $ \Gamma=-\Im \Sigma $ for simplicity, $h$ is the Planck constant, and the degeneracies $ g_s=g_v=2 $ have been considered. As for MLG, the longitudinal conductivity has a similar expression $ \sigma^0_{xx} =\frac{2e^2}{\pi h} [1+(\frac{\alpha}{\Gamma} +\frac{\Gamma}{\alpha}) \arctan \frac{\alpha}{\Gamma}] $. Obviously, the minimum conductivity of BLG calculated by the two-band model is a universal value $ \sigma^0_{\text{min}}= \frac{8e^2}{\pi h} $ and is twice as large as that of MLG ($ \sigma^0_{\text{min}} =\frac{4e^2}{\pi h} $). Based solely on the minimum conductivity at zero temperature, the BLG appears to be a simple superposition of two MLG layers.

	\subsection{Interlayer Coupling}
	\label{sec.inter}
	\begin{figure}[t!]
		\centering
		\includegraphics[width=1 \linewidth]{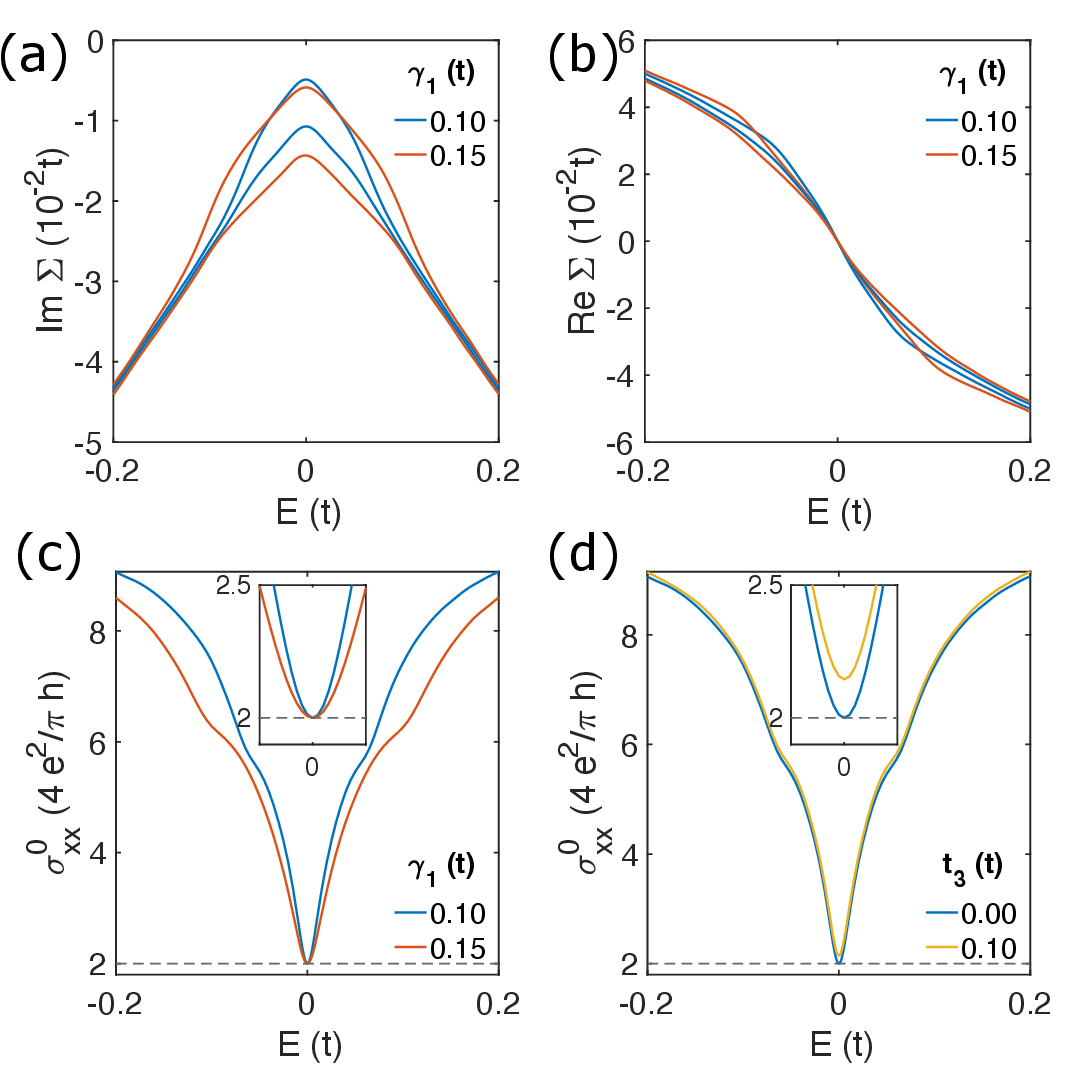}
		\caption{(a) Imaginary and (b) real parts of the self-energy as functions of energy for bilayer graphene with different interlayer coupling strengths $ \gamma_1=0.10t,0.15t $. (c) Conductivity $ \sigma_{xx} $ as a function of energy with different interlayer coupling strengths $ \gamma_1=0.10t,0.15t $. (d) Conductivity $ \sigma_{xx} $ with different interlayer coupling strengths $ t_3=0,0.10t $ and $ \gamma_1=0.1t $. The insets in (c) and (d) show the conductivity within the energy window $ [-0.01t,0.01t] $. Here, the disorder strength $ u_0=0.09 $.}
		\label{fig.inter}
	\end{figure}
	
	To investigate the role of interlayer coupling $ \gamma_1 $, we calculated self-energy $ \Sigma $ and zeros temperature conductivity $ \sigma^0_{xx} $ with different $ \gamma_1 $, as shown in Fig.~\ref{fig.inter}(a)(b)(c), and the differences are clear. The split of $ \Sigma^{\text{pw}}_1 $ and $ \Sigma^{\text{pw}}_2 $ become more obvious and the conductivity decreases as $ \gamma_1 $ increases. It is worth noting that $ \gamma_1 $ has no effect on minimum conductivity at zero temperature, but when the temperature is not zero, the minimum conductivity decreases with increasing $ \gamma_1 $.
	
	In a more realistic scenario, there will be some interlayer jumps other than $ \gamma_1 $, such as the interlayer interactions between $ \text{A}_1 $ and $ \text{B}_2 $, denoted as $ t_3 $ (or called trigonal warping term), and the corresponding Hamiltonian is written as
	\begin{align*}
		H_{w}=-t_3\sum_{\langle i,j\rangle} c^{\dag}_{\text{A}_1,i} c_{\text{B}_2,j} +\text{h.c.}
	\end{align*}
	As shown in the Fig.~\ref{fig.inter}(d), when $ t_3\neq 0 $, the minimum conductivity at zero temperature $ \sigma_{\text{min}}^0 $ no longer be a universal value, but will become larger. This is because this type of interlayer terms is non-local and can also be reflected in the velocity operator.
	
	\subsection{Higher Order Conductivity Corrections}
	Based on the bare current bubble which contributes the most to the classical conductivity, we also consider the vertex correction and the quantum interference correction, whose corresponding diagrams are the ladder diagram and the maximally-crossed diagram, respectively. 
	
	Under the ladder approximation, the renormalized velocity $ \tilde{v}_x $ satisfies the Beta-Salpeter (BS) equation \cite{shon1998quantum}
	\begin{align}
		&\tilde{v}_x(\bm{k}, E+i\eta,E-i\eta) =v_x +\sum_{\bm{k}^\prime} \langle V(\bm{k} -\bm{k}^\prime) G^R(\bm{k}^\prime,E) \nonumber
		\\ &\times \tilde{v}_x (\bm{k}^\prime, E+i\eta,E-i\eta) G^A(\bm{k}^\prime,E) V(\bm{k}^\prime -\bm{k}) \rangle,
	\end{align}
	and we assume that the renormalized velocity $ \tilde{v}_x=\Lambda v_x $ differs from the bare velocity only by an energy dependent constant $ \Lambda $. For the short-range disorder considered here, the current operator only has a correction on the diagonal term. Because the velocity operator in BLG is non-diagonal, the vertex correction vanishes here.
	
	The quantum interference correction to the conductivity is associated with disorder-averaged Cooperon function, and the derivation process is shown in the Appendix \ref{app.correcction}. In the absence of trigonal warping, the only Cooperon channel that remains gapless is the sublattice-triplet and valley-triplet Cooperon, which belongs to the intervalley channel category. The quantum interference conductivity correction is obtained to be negative (corresponding to weak localization)  and can be evaluated as
	 \begin{align}
			\sigma_{\text{qi}}\simeq-\frac{e^{2}}{\pi h}\ln\frac{\min\{L,L_{\phi}\}}{\ell_{e}},
		\end{align}
		where $ L $ is the length of system, $ L_{\phi} $ is the coherence length, and $ \ell_e(E,\Sigma) \propto n_{\text{imp}} u^2 $ is the mean free path. Although BLG, like MLG, is chiral (with the additional quantum number, pseudospin, originating from the sublattice freedom), there is no backscattering suppression because it has a $2\pi$ berry phase rather than $\pi$ indicating the quantum correction is conventional weak localization. When the trigonal warping term is taken into  consideration, the quantum interference will be suppressed, unless the intervalley scattering is sufficiently strong \cite{mccann2013electronic,mccann2006weaklocalization,kechedzhi2007influence,gorbachev2007weak}. According to the scaling theory, the scaling function $\beta(g)=d(\ln g)/d(\ln L)$ is negative, then conductivity $g$ decreases as the system size $L$ is enlarged and the system is insulating in the thermodynamic limit. The inclusion of higher-loop corrections may provide a more complete understanding of the localization effects, worthy for future studies. 
	
	\subsection{Comparison with Experiments}
	
	\begin{figure}[t!]
		\centering
		\includegraphics[width=1 \linewidth]{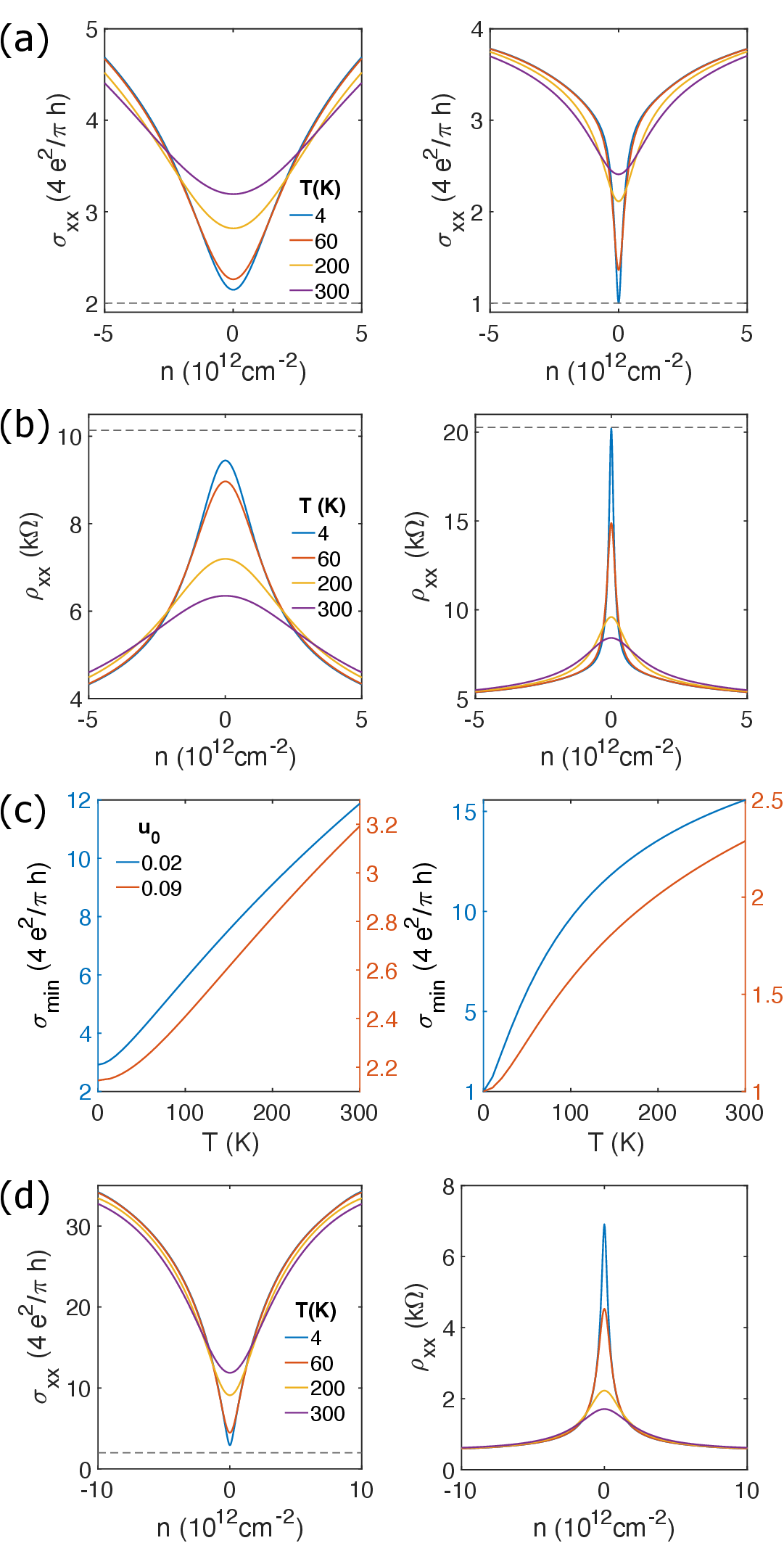}
		\caption{(a) Conductivity $ \sigma_{xx} $ and (b) resistivity $ \rho_{xx} $ as functions of the carrier density $ n $ at different temperatures $ T=4,60,200,\SI{300}{K} $, with $ u_0=0.09 $. (c) Temperature dependence of the minimum conductivity $ \sigma_{\text{min}} $ with $ u_0=0.02,0.09$. In (a-c), the left and right panels are for BLG and MLG, respectively. (d) Conductivity and resistivity for BLG, with $u_0=0.02$. Here, the trigonal warping $t_3=0.1t$.}
		\label{fig.cond}
	\end{figure}
	
	For better comparison with experiments and with MLG, the conductivity $ \sigma_{xx} $ and resistivity  $ \rho_{xx} =1/\sigma_{xx} $ as a function of carrier density $ n $ (corresponds to the gate voltage in experiments) at different temperatures $ T =4, 60, 200,300\ \si{K} $ are plotted in Figs.~\ref{fig.cond}(a) and ~\ref{fig.cond}(b), respectively, for both BLG and MLG. Here, the carrier density is given by $ n=\int_0^{\infty} D(E) f(E,E_f)dE +\int_{-\infty}^0 D(E) [1-f(E,E_f)]dE $, with $ D(E) $ denoting the DOS, and a typical weak disorder strength is chosen as $ u_0 = 0.09 $. It is found that the conductivity of MLG is strongly sublinear with varying the carrier density, indicating that the similar experimentally sublinear behavior can occur with only short-range impurities. We also observe that the conductivity (resistivity) of both systems has a strong temperature dependence, with a sharp dip (peak) emerging at low temperature. And there exists a critical carrier density $ n^* $ that divides the carrier density into two regimes. In the low doping regime, the conductivity shows a strong insulating temperature dependence, that is, the conductivity decreases with decreasing temperature ($ d\sigma/d T>0 $).  At high carrier density, such as $ n=5\times 10^{12}\ \si{cm}^{-2} $ for BLG ($ n=3\times 10^{12}\ \si{cm}^{-2} $ for MLG), the conductivity exhibits a weak temperature dependence with a metallic temperature behavior $ d\sigma/d T<0 $.  The critical carrier density of BLG is much larger than the value of MLG. For the same disorder strength parameter, the conductivity and resistivity curves of BLG are less acute and more rounded at the CNP than those of MLG. The conductivity and resistivity of the BLG with $u_0=0.02$ are also drawn in Fig.~\ref{fig.cond}(d). The $\sigma(n)$ is sublinear, and consistent with the observation in the suspended sample \cite{feldman2009brokensymmetry}. The main features of the temperature and carrier dependence of our calculated results, such as the sublinear behavior and the existence of a critical carrier density dividing insulating and metallic temperature dependences, are in agreement with the experiments \cite{du2008approaching,bolotin2008temperature,dean2010boron,zomer2011transfer,mayorov2011micrometerscale,mayorov2011interactiondriven,mayorov2012close,amin2018exotic,defazio2019highmobility}.
	Moreover, Fig.~\ref{fig.cond}(c) depicts the minimum conductivity of both systems increasing monotonically with temperature, but their slopes are very different.
	The minimum conductivity of MLG varies linearly at first, then sublinearly with temperature; in contrast, that of BLG is parabolic at low temperature and linear at higher temperature.
	The minimum conductivity at zero temperature for MLG is universal $\frac{4e^2}{\pi h}$, whereas for BLG, it deviates from the universal value of $\frac{8e^2}{\pi h}$ (indicated by the dotted line in Fig.~\ref{fig.cond}(d)) due to the trigonal warping term  and is correlated with the disorder strength. The overall trend of the temperature  dependency of the minimum conductivity agrees with experiments, particularly for MLG, but the linear behavior for BLG is a bit different from that in experiments, probably because there are more parameters which are out of our consideration \cite{dean2010boron,du2008approaching,feldman2009brokensymmetry,mayorov2011interactiondriven,mayorov2011micrometerscale,nam2017electron}.

	\section{Conclusion}
	\label{sec.discussion}

	In summary, we have studied the quasiparticle and transport properties of bilayer graphene with short-range Anderson disorder using the Lanczos method. The quasiparticle properties have been investigated in both the strong and weak scattering limits, revealing that the quasiparticle residue decreases significantly near the charge neutrality point. These intriguing features are the consequences of multiple impurity scattering events, which can be captured by using the Lanczos method. Furthermore, we found that the conductivity increases with the carrier density and saturates at high carrier densities, and the interlayer scattering events will reduce the longitudinal conductivity. We also obtained the characteristic dependence of the conductivity on the temperature in the low carrier density limit. The results have been further compared with the findings of the same system using the self-consistent Born approximation, demonstrating the pronounced differences between the two approaches, as well as the central findings in the preceding paper for monolayer graphene, highlighting significant interlayer scattering effects. Noticeably, at low carrier density, the conductivities of BLG and MLG exhibit parabolic and sublinear behaviors, respectively, and both have a critical carrier density that separates the strong insulating and weak metallic regimes characterized by the temperature dependence of resistivity. The overall trends of our numerical results are in good agreement with experimental observations. Furthermore, in the absence of trigonal warping term, the minimum conductivity of BLG at zero temperature is a universal constant of $ 8e^2/\pi h $, and is independent of the interlayer scattering strength $ \gamma_1 $. But when trigonal warping term is considered, the minimum conductivity of BLG at zero temperature is dependent on disorder strength.
	Our findings help to provide some new angles into the quasiparticle and transport properties of disordered bilayer graphene.

	\begin{acknowledgments}
		This work is supported by the National Natural Science Foundation of China (Grant No. 11974323), the Innovation Program for Quantum Science and Technology (Grant No. 2021ZD0302800), the Anhui Initiative in Quantum Information Technologies (Grant No. AHY170000), and the Strategic Priority Research Program of Chinese Academy of Sciences (Grant No. XDB30000000).
	\end{acknowledgments}

	\appendix
	
	\section{Formulation}
	By taking K $_+$ valley ($ \xi= + $) for example,
	\begin{align}
		& H_0(\theta)= \begin{pmatrix}
			0 & \hbar v_f k_- & 0 & 0 \\
			\hbar v_f k_+ & 0 & \gamma_1 & 0 \\
			0 & \gamma_1 & 0 & \hbar v_f k_- \\
			0 & 0 & \hbar v_f k_+ & 0
		\end{pmatrix},
	\end{align}
	where $ k_{\pm}= k_x\pm ik_y $. 
	The eigenvalues are obtained as
	\begin{align}
		& \varepsilon(\bm{k}) =\pm \left[ \sqrt{ (\hbar v_f k)^2 + \left(\frac{\gamma_1}{2} \right)^2} \pm \frac{\gamma_1}{2} \right],
	\end{align}
	with the eigenstates
	\begin{align}
		\Psi (\theta)= \frac{1}{\sqrt{\mathcal{S}}}\ \varphi(\theta)\ e^{i\bm{k} \cdot \bm{r}},
	\end{align}
	where $ k=|\bm{k}| =\sqrt{k_x^2 +k_y^2} $, $ \theta =\atan (k_y/k_x) $, and $ \mathcal{S} $ is the area of the system.
	
	To obtain a concise analytical solution, we introduce a transform matrix  \cite{ando2011bilayer,ando2015theory,ando2019theory} 
	\begin{align}
		U(\theta)=
		\begin{pmatrix}
			1 & 0 & 0 & 0 \\
			0 & e^{i\theta} & 0 & 0 \\
			0 & 0 & e^{i\theta} & 0 \\
			0 & 0 & 0 & e^{2i\theta}
		\end{pmatrix}
	\end{align}
	to eliminate the angular dependence on the direction of $ \bm{k} $.
	The Hamiltonian becomes
	\begin{align}
		\begin{split}
			\tilde{H}_0 &=U^{-1}(\theta) H_0(\theta) U(\theta)
			\\& =
			\begin{pmatrix}
				0 & \hbar v_f k & 0 & 0 \\
				\hbar v_f k & 0 & \gamma_1 & 0 \\
				0 & \gamma_1 & 0 & \hbar v_f k \\
				0 & 0 & \hbar v_f k & 0
			\end{pmatrix},
		\end{split}
	\end{align}
	and therefore the corresponding eigenstates are given by
	\begin{align}
		&\tilde{\varphi} =U^{-1}(\theta)\ \varphi(\theta),
		\\& \Psi (\theta)= \frac{1}{\sqrt{\mathcal{S}}}\ U(\theta)\ \tilde{\varphi}\ e^{i\bm{k} \cdot \bm{r}}.
	\end{align}
	As shown above, the angular dependence $ \theta=\theta(\bm{k}) $ is absorbed into the matrix $ U(\theta) $.
	
	Furthermore, we can define
	\begin{align}
		&\epsilon(k)=\sqrt{(\hbar v_f k)^2 + (\frac{\gamma_1}{2})^2},
		\\ & \frac{\gamma_1}{2} = \epsilon(k) \cos \psi,
		\\ & \hbar v_f k=  \epsilon(k) \sin \psi.
	\end{align}
	Additionally, $ \psi =\atan(\frac{\hbar v_fk}{\gamma_1/2}) $, with $ \psi=0 $ for $ k=0 $ and $ \psi=\frac{\pi}{2} $ for $ k\rightarrow \infty $. The eigenvalues are rewritten as
	\begin{align}
		\begin{split}
			\varepsilon_{\pm 1}
			&=\pm \left[ \sqrt{(\hbar v_f k)^2 + (\frac{\gamma_1}{2})^2} + \frac{\gamma_1}{2} \right]
			\\&= \pm 2\epsilon(k) \cos^2\frac{\psi}{2}	,
		\end{split} 
	\end{align}
	\begin{align}
		\begin{split}
			\varepsilon_{\pm 2} 
			&=\pm \left[ \sqrt{(\hbar v_f k)^2 + (\frac{\gamma_1}{2})^2} - \frac{\gamma_1}{2} \right]
			\\ &= \pm 2\epsilon(k) \sin^2\frac{\psi}{2},
		\end{split}
	\end{align}
	where the subscripts of $ ``1"$ and $``2" $ represent different band indexes, and the $``+" $ and $``-"$ represent the conduction and valence bands, respectively.
	Therefore, we can easily get
	\begin{align}
		S= \frac{1}{\sqrt{2}} \begin{pmatrix}
			-\sin \frac{\psi}{2} & -\cos \frac{\psi}{2} & \cos \frac{\psi}{2} & \sin \frac{\psi}{2} \\
			\cos \frac{\psi}{2} & \sin \frac{\psi}{2} & \sin \frac{\psi}{2} & \cos \frac{\psi}{2} \\
			-\cos \frac{\psi}{2} & \sin \frac{\psi}{2} & -\sin \frac{\psi}{2} & \cos \frac{\psi}{2} \\
			\sin \frac{\psi}{2} & -\cos \frac{\psi}{2} & -\cos \frac{\psi}{2} & \sin \frac{\psi}{2}
		\end{pmatrix}.
	\end{align}
	where the columns are the corresponding eigenvectors $ \tilde{\varphi} $ for $ \varepsilon_{-1} $, $ \varepsilon_{-2} $, $ \varepsilon_{2} $, $ \varepsilon_{1} $ (or labeled as $ \varepsilon_n $, $ n=1,2,3,4 $), ordered from the lowest to highest bands, respectively.

	\section{Self-consistent Born approximation (SCBA)}
	\label{sec.SCBA}

	By considering the short-range disorder whose potential range is much smaller than the lattice constant, the random on-site potential is given as 
	\begin{align}
		\begin{split}
			V (\bm{r}) &= \sum_{\bm{r}} \sum_{\mathcal{A}} u_{\mathcal{A}} (\bm{r}) c^{\dag}_{\bm{r}, \mathcal{A}} c_{\bm{r}, \mathcal{A}}
			\\ &=\sum_b \Gamma_b \left(\sum_i u_i^b \delta(\bm{r} -\bm{R}_i) \right)
			\\ &=\sum_{b=1}^4 \Gamma_b V_{\bm{r}}^b
		\end{split}.
	\end{align}
	Where, $ \mathcal{A}= \text{A}_1, \text{B}_1, \text{A}_2, \text{B}_2 $, the $ b $-th diagonal element of the matrix $ \Gamma_b $ is $ u_i^b $ and all the others are zero. The average over the impurity configuration of the potential is $ \langle V_{\bm{r}}\rangle=0 $, and the potential correlation is 
	\begin{align}
		\begin{split}
			\langle V_{\bm{r}} \otimes V_{\bm{r}^\prime} \rangle &=\sum_{bb^\prime}  \langle V_{\bm{r}}^b V_{\bm{r}^\prime}^{b^\prime} \rangle \Gamma_b \otimes \Gamma_{b^\prime}
			\\ &= \sum_{bb^\prime} \langle | u_i^b|^2\rangle \delta_{bb^\prime} \delta(\bm{r} -\bm{r}^\prime)  \Gamma_b \otimes \Gamma_{b^{\prime}}
			\\ &= n_{\text{imp}} u^2  \delta(\bm{r} -\bm{r}^\prime)  \sum_b \Gamma_b \otimes \Gamma_{b}
		\end{split}.
	\end{align}
	Here, $ \langle |u_i^b|^2 \rangle = n_{\text{imp}} u^2 $ and $ u^2=  A_c \frac{ W^2}{12} $ with no sublattice disorder correlation, and $ \langle \cdots\rangle $ means the disorder averaging.
	
	\subsection{Eigenstate Representation}
	To be able to compare directly with the results calculated by the Lanczos method, we also calculate the self-energy of clear BLG in the eigenstate representation. The potential is rewritten as
	\begin{align}
		V_{\bm{k}- \bm{k}^\prime} =\int d^2 \bm{r}\ e^{-i (\bm{k} -\bm{k}^\prime) \cdot \bm{r}} (US)_{\bm{k}}^\dag V_{\bm{r}} (US)_{\bm{k}^\prime}.
	\end{align}
	With the Born approximation (BA), we can define the self-energy as the follow form
	\begin{align}
		\Sigma(\bm{k},E)=\begin{pmatrix}
			\Sigma_1 & 0 & \Sigma_3 & 0 \\
			0 & \Sigma_2 & 0 & \Sigma_3 \\
			\Sigma_3 & 0 & \Sigma_2 & 0 \\
			0 & \Sigma_3 & 0 & \Sigma_1
		\end{pmatrix}, 
	\end{align}
	and the Green's function can be written as
	\begin{align}
		G=\begin{pmatrix}
			G_1 & 0 & G_{31} & 0 \\
			0 & G_2 & 0 & 0 \\
			G_{31} & 0 & G_3 & G_{42} \\
			0 & G_{42} & 0 & G_4
		\end{pmatrix}.
	\end{align}
	Then, we can get
	\begin{align}
		\begin{split}
			\Sigma(k,E) &=  \int \frac{d^2 \bm{k}^\prime}{(2\pi)^2}
			\langle V_{\bm{k}- \bm{k}^\prime} G(\bm{k}^\prime, E) V_{\bm{k}^\prime -\bm{k}} \rangle
			\\ &= \frac{n_{\text{imp}} u^2}{2} \int \frac{d^2 \bm{k}^\prime}{(2\pi)^2}
			\begin{pmatrix}
				C & 0 & F & 0  \\
				0 & D & 0 & F \\
				F & 0  & D & 0 \\
				0 & F& 0 & C
			\end{pmatrix},
		\end{split}
	\end{align}
	with the matrix elements
	\begin{align}
		\begin{split}
			C&=g1+g3\cos\psi_k,
			\\D&=g1-g3\cos\psi_k,
			\\F&= g3\sin\psi_k.
		\end{split}
	\end{align}
	Here,
	\begin{align}
		\label{g1g2}
		\begin{split}
			&g1=G_1+G_2+G_3+G_4,
			\\ &g2=G_1-G_2-G_3+G_4,
			\\& g3=g2 \cos\psi_{k^\prime}+ 2(G_{31}+G_{42}) \sin\psi_{k^\prime}.
		\end{split}
	\end{align}
	
	From the above derivations, we find that the self-energy is not diagonal and has a momentum dependence.
	
	\subsection{Plane Wave Representation}
	By Fourier transform, the momentum-space matrix elements of the disorder potential are given as
	\begin{align}
		V_{\bm{k}- \bm{k}^\prime} =\int d^2 \bm{r}\ V_{\bm{r}}\ e^{-i (\bm{k} -\bm{k}^\prime) \cdot \bm{r}} .
	\end{align}
	The self-energy calculated by the SCBA is
	\begin{widetext}
		\begin{align}
			\begin{split}
				\Sigma^{\text{pw}}(\bm{k}, E) &= \int \frac{d^2 \bm{k}^\prime}{(2\pi)^2} \langle V_{\bm{k}- \bm{k}^\prime} G(\bm{k}^\prime, E) V_{\bm{k}^\prime -\bm{k}} \rangle
				\\&=n_{\text{imp}} u^2 \sum_{b=1}^4   \int_0^{k_c} \frac{d^2 \bm{k}^\prime}{(2\pi)^2} \Gamma_b G(\bm{k}^\prime, E) \Gamma_b
				\\& = \frac{n_{\text{imp}} u^2}{ (\hbar v_f)^2} \int_0^{E_c} \frac{x dx}{2\pi} 
				\frac{2}{[(x^2-\omega_1\omega_2)^2 -(\omega_1\gamma_1)]} 
				\\ &\qquad \times \begin{pmatrix}
					-\gamma_1^2\omega_1 -\omega_2(x^2-\omega_1\omega_2) & 0 & 0 & 0 \\
					0 & -\omega_1(x^2-\omega_1\omega_2) & 0 & 0 \\
					0 & 0 & -\omega_1(x^2-\omega_1\omega_2) & 0 \\
					0 & 0 & 0 & -\gamma_1^2\omega_1 -\omega_2(x^2-\omega_1\omega_2)
				\end{pmatrix}
				\\&=\begin{pmatrix}
					\Sigma_1^{\text{pw}} & 0 & 0 & 0 \\
					0 & \Sigma_2^{\text{pw}}  & 0 & 0 \\
					0 & 0 & \Sigma_2^{\text{pw}}  & 0 \\
					0 & 0 & 0 & \Sigma_1^{\text{pw}} 
				\end{pmatrix}
				\\ &\propto \frac{n_{\text{imp}} u^2}{\pi (\hbar v_f)^2}
			\end{split}.
		\end{align}
	\end{widetext}
	Therefore, we define a dimensionless disorder strength 
	\begin{align}
		u_0=\frac{n_{\text{imp}} u^2}{ \pi (\hbar v_f)^2} =\frac{n_{\text{imp}} A_c W^2}{ 12 (\hbar v_f)^2 \pi} .
	\end{align} 
	Here, 
	$ x=\hbar v_f k $ is the rescaling integration variable, and $ E_c=\hbar v_f k_c $ is the high energy cutoff with $ E_c=2.7t  $. In the BA, $ \omega_1=\omega_2=E+i\eta $, and in the SCBA, $ \omega_{1/2} =E-\Sigma_{1/2} $. The self-energy is diagonal under the action of $ \Gamma_b $. After integration,
	\begin{widetext}
		\begin{align}
			\begin{split}
				\Sigma^{\text{pw}}_1 (E)&= -\frac{u_0}{4} \left\{ \gamma_1 \ln \frac{(E_c^2-\omega_1\omega_2 -\omega_1\gamma_1)(-\omega_1\omega_2 +\omega_1\gamma_1)} {(E_c^2-\omega_1\omega_2 +\omega_1\gamma_1) (-\omega_1\omega_2 -\omega_1\gamma_1)} 
				+\omega_2\ln \frac{(E_c^2-\omega_1\omega_2)^2- (\gamma_1\omega_1)^2}{(\omega_1\omega_2)^2- (\gamma_1\omega_1)^2}  \right\}
				\\ &\approx -i\frac{\pi \gamma_1}{4} u_0 -\frac{u_0(E-\Sigma_2^{\text{pw}})}{4}  \ln \frac{E_c^4}{-\left[ \gamma_1(E-\Sigma_1^{\text{pw}}) \right]^2}, 
			\end{split}
		\end{align}
		\begin{align}
			\begin{split}
				\Sigma^{\text{pw}}_2 (E)&=-\frac{u_0\omega_1}{4} \ln \frac{(E_c^2-\omega_1\omega_2)^2- (\gamma_1\omega_1)^2}{(\omega_1\omega_2)^2- (\gamma_1\omega_1)^2} 
				\\& \approx -\frac{u_0(E-\Sigma_1^{\text{pw}})}{4}  \ln \frac{E_c^4}{- \left[ \gamma_1(E-\Sigma_1^{\text{pw}}) \right]^2} .
			\end{split}
		\end{align}
	\end{widetext}
	The self-consistent equation can be solved numerically by iteration. Apparently, the self-energy is $ \bm{k} $ independent. The intra- and inter-valley scattering processes contribute equally to the self-energy here \cite{shon1998quantum}, and have both been considered in the above equation . 
	
	
	When $ \gamma_1=0 $,
	\begin{align}
		\Sigma_1=\Sigma_2=\Sigma\approx -u_0 (E-\Sigma)\ln \frac{E_c}{E-\Sigma}.
	\end{align}
	When $ E=0 $, due to the electron-hole symmetry, the real part of the self-energy is zero, and we can define $ \Sigma(E=0)=-i\Gamma(0)$ and
	\begin{align}
		\Gamma(0)=E_c e^{-1/u_0}.
	\end{align}

	\subsection{Representation Transformation}
	\label{sec.transform}
	The transformation of the self-energy from the plane wave representation to eigenstate representation are given as
	\begin{align}
		\begin{split}
			\Sigma_1 &= \frac{1}{2} \left[ (\Sigma_1^{\text{pw}}+\Sigma_2^{\text{pw}})- (\Sigma_1^{\text{pw}} -\Sigma_2^{\text{pw}})\cos\psi_k \right],
			\\ \Sigma_2 &= \frac{1}{2} \left[ (\Sigma_1^{\text{pw}}+\Sigma_2^{\text{pw}})+ (\Sigma_1^{\text{pw}} -\Sigma_2^{\text{pw}})\cos\psi_k \right],
			\\ \Sigma_3&= -\frac{1}{2}  (\Sigma_1^{\text{pw}}-\Sigma_2^{\text{pw}}) \sin\psi_k.
		\end{split}
	\end{align}

	\subsection{Self-energy at $ E=0 $}

	In order to investigate more intuitively the self-energy at the CNP, we can adopt the two-band model and obtain
	\begin{align}
		\begin{split}
			\Sigma(E) =&n_{\text{imp}} u^2 \int \frac{k^
				\prime dk^\prime}{2\pi} \left( \frac{1}{E-\Sigma -x} 
			+ \frac{1}{E-\Sigma +x} \right) I_{2\times 2},
		\end{split}
	\end{align}
	where $ x=k^{\prime 2}/2m $ with $ m=\gamma_1/2 (\hbar v_f)^2 $.
	Then, $ \Gamma(0)=-\Im \Sigma(E=0) $ can be easily given as 
	\begin{align}
		\begin{split}
			\Gamma(0) 
			&=\frac{n_{\text{imp}} u^2 m}{\pi} \Gamma(0)\int dx  \frac{1}{x^2+\Gamma(0)^2}  
			\\&=\frac{n_{\text{imp}} u^2 m}{\pi} \arctan \frac{E_c}{\Gamma(0)}
			\\ &\approx \frac{ n_{\text{imp}} u^2 m}{2} 
			\\&=\frac{\pi\gamma_1}{4} u_0
		\end{split}.
	\end{align}
	The $ \Gamma(0) $ calculated by the Lanczos method, four-band model with the SCBA, and two-band model with the BA are compared in Fig.~\ref{fig.ImS0}. The three methods clearly fit better when the disorder strength $ u_0 $ is weak, and the BA and SCBA cannot encompass all of the disorder effects well when the disorder strength increases.

	\begin{figure}[t!]
		\centering
		\includegraphics[width=1\linewidth]{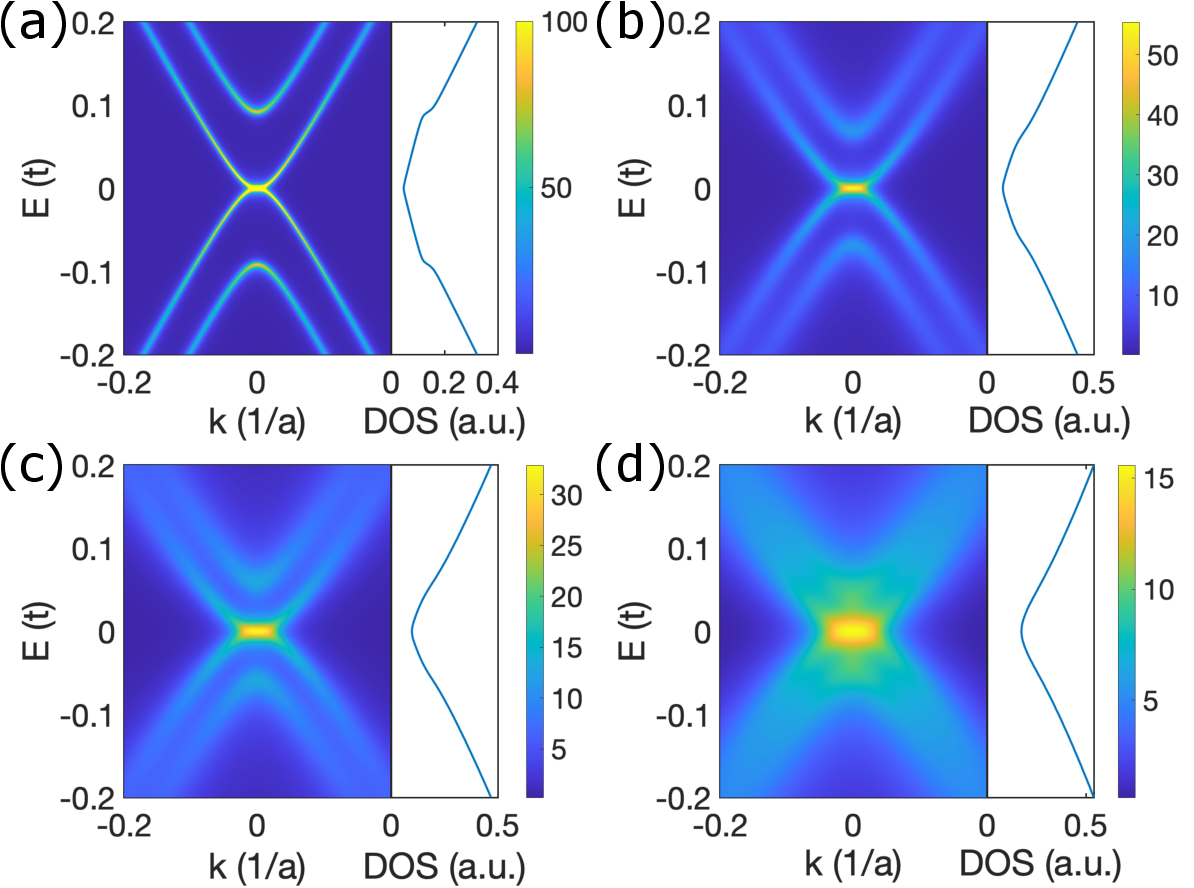}
		\caption{Spectral function $ A(\bm{k},E) $ and density of states (DOS) with different disorder strengths $ u_0 $. (a) $ u_0=0.02 $, (b) $ u_0=0.09 $, (c) $ u_0=0.12 $, and (d) $ u_0=0.17 $.}
		\label{fig.spectral_u0}
	\end{figure}
	
	\section{Spectral Function and Density of States}
	\label{app.spectral}

	The single-particle spectral function is related to the Green's function by the relation of
	
	\begin{align}
		\begin{split}
			A_n(\bm{k},E) &=-\frac{1}{\pi} \Im G_n(\bm{k},E)
			\\& =\frac{1}{\pi} \frac{-\Im \Sigma}{\left[ E-\varepsilon_n(\bm{k}) -\Re \Sigma(\bm{k},E) \right]^2 +\left[-\Im \Sigma(\bm{k},E)\right]^2}
			\\& =\frac{1}{\pi} \frac{\Gamma} {(\alpha-\varepsilon_n)^2 +\Gamma^2}
		\end{split},
	\end{align}
	with $ \alpha=E-\Re \Sigma(\bm{k},E) $ and $ \Gamma=-\Im \Sigma(\bm{k},E) $ for simplicity. The spectral function $ A(\bm{k},E) $ is a $ \delta $ function in the absence of the disorder, indicating that the wave vector is a good quantum number. The spectral function is broadened and the quasiparticles have a finite lifetime when the disorder is introduced, as shown in Fig.~\ref{fig.quasi}(b) and Fig.~\ref{fig.spectral_u0}. The dispersion relation is represented by the peak of the spectral function $ A(\bm{k},E) $. 
	As the disorder strength increases, the peak of $ A(\bm{k},E) $ moves toward $ E = 0 $, indicating that the dispersion relation is strongly renormalized due to the multiple scattering.
	
	The DOS per unit cell can also be obtained based on the spectral function in the momentum space as
	\begin{align}
		D(E) &= g_sg_v \frac{1}{N} \sum_{\bm{k}} \Tr A(\bm{k},E)
		\\& = A_c g_s g_v \int\frac{d^2 \bm{k}}{(2\pi)^2}  \sum_{n=1}^4 A_n(\bm{k},E).
	\end{align}
	The corresponding results plotted in Fig.~\ref{fig.spectral_u0} and Fig.~\ref{fig.quasi}(b) agree well with the real-space results plotted in Fig.~\ref{fig.quasi}(a) by the Lanczos method.

	\section{Conductivity Correction}
	\label{app.correcction}
	
	\begin{figure}
	\includegraphics[width=1\linewidth]{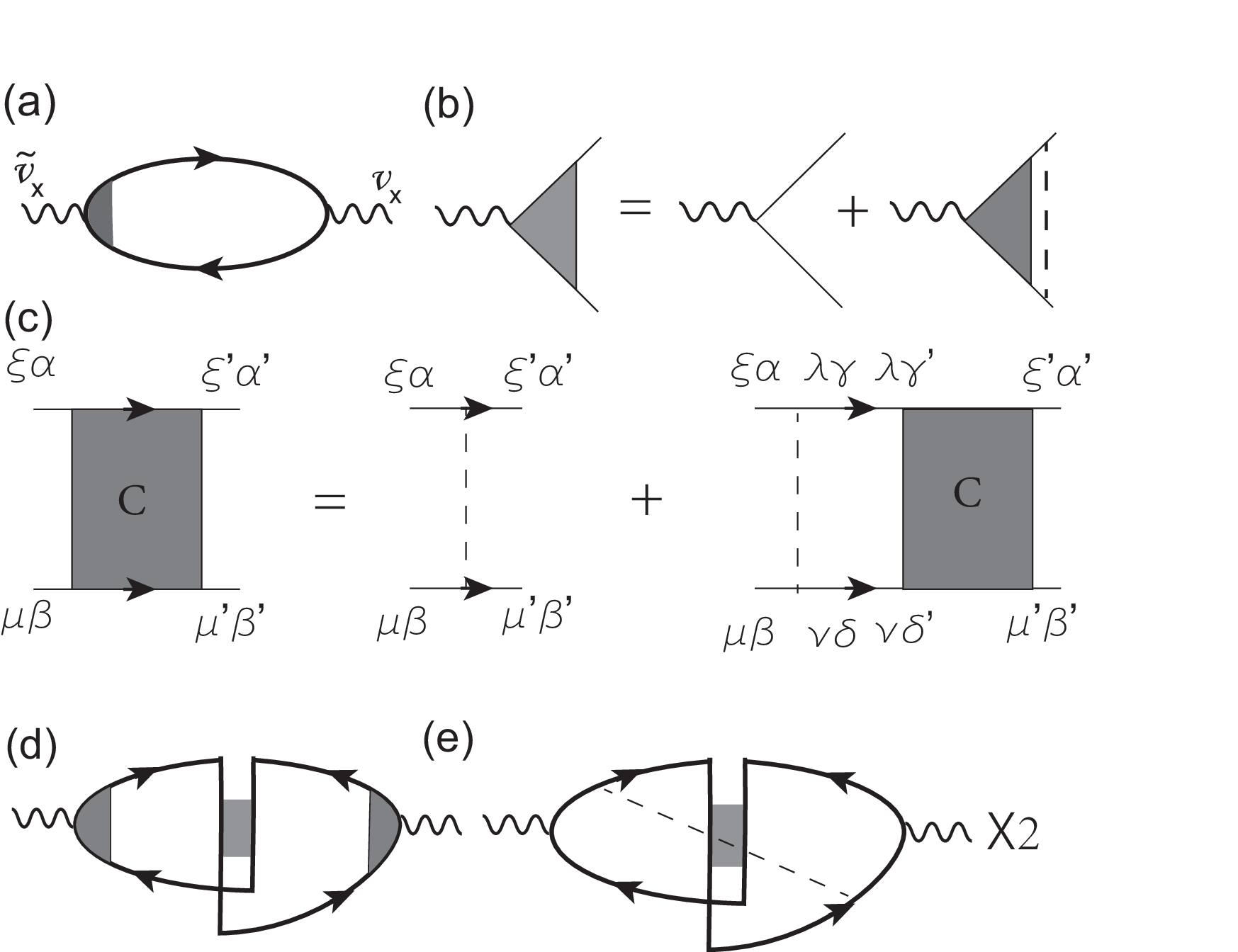}
	\caption{(a) Diagram for the Drude conductivity (b) The vertex correction. (c) Bethe-Salpeter equation for the Cooperon propagator. (d) Bare Hikami box for quantum interference correction (e) The dressed Hikami boxes. Solid lines represent disorder averaged Green's function and dashed lines represent disorder.\label{fig:Feynman}}
\end{figure}

In this part, we derive the quantum interference correction to the conductivity
of bilayer graphene for a short-range, uncorrelated disorder potential following
the references \cite{mccann2006weaklocalization,kechedzhi2007influence}. The low-energy spectrum of
Bernal-stacked bilayer graphene can be described by the Hamiltonian 
\begin{equation}
H=\frac{1}{2m}\varPi_{0}[(p_{x}^{2}-p_{y}^{2})\sigma_{x}+2p_{x}p_{y}\sigma_{y}]\label{eq:Hamiltonian},
\end{equation}
which acts in the space of four-component wave functions $\Phi=[\phi_{\mathbf{K}_{+},A_{1}},\phi_{\mathbf{K}_{+},B_{2}},\phi_{\mathbf{K}_{-},B_{2}},\phi_{\mathbf{K}_{-},A_{1}}]$.
Here the Pauli matrices $\boldsymbol{\sigma}$ and $\boldsymbol{\varPi}$ act in
sublattice and valley spaces respectively. The current operators from Eq. (\ref{eq:Hamiltonian})
are momentum dependent $\boldsymbol{v}=\frac{\partial H}{\partial \bm{p}}= \frac{1}{m}(p_{x}\sigma_{x}+p_{y}\sigma_{y},-p_{y}\sigma_{x}+p_{x}\sigma_{y}).$
The disorder Hamiltonian takes the form 
\begin{align}
\hat{V}_{\text{dis}}=\sum_{\mathbf{r}}\left(u_{A_{1}}(\boldsymbol{r})\phi_{\boldsymbol{r},A_{1}}^{\dagger}\phi_{\boldsymbol{r},A_{1}}+u_{B_{1}}(\boldsymbol{r})\phi_{\boldsymbol{r},B_{2}}^{\dagger}\phi_{\boldsymbol{r},B_{2}}\right),
\end{align}
where different types of disorder are uncorrelated on averaging $\langle u_{\mathcal{A}}(\boldsymbol{r})u_{\mathcal{A}^{\prime}}(\boldsymbol{r}^{\prime})\rangle=nu^{2}\delta_{\mathcal{A}\mathcal{A}^{\prime}}\delta(\boldsymbol{r}-\boldsymbol{r}^{\prime})$. In
momentum space, the scattering amplitude is $u_{\mathcal{A}}(\boldsymbol{k})=\frac{1}{N}\sum_{\boldsymbol{r}}u_{\mathcal{A}}(\boldsymbol{r})e^{i\boldsymbol{k}\cdot\boldsymbol{r}}$.
By expressing $\boldsymbol{k}=\bm{q}+(\mathbf{K}_{\xi}-\mathbf{K}_{\xi^{\prime}})$
in terms of a small momentum transfer $\boldsymbol{q}$ within a single valley and
the scattering amplitude can be separated into an intravalley part ($\xi=\xi^{\prime}$)
and an intervalley part ($\xi\ne\xi^{\prime}$). Then the disorder Hamiltonian in
the four-component basis $\Phi$ is
\begin{align}
	\hat{V}_{\text{dis}}\simeq\sum_{\boldsymbol{p},\boldsymbol{p}^{\prime}}\Phi_{\boldsymbol{p}}^{\dagger}V_{\boldsymbol{p}-\boldsymbol{p}^{\prime}}\Phi_{\boldsymbol{p}^{\prime}},
\end{align}
with 
\begin{align}
V_{\boldsymbol{q}}=\left(\begin{array}{cccc}
u_{A_{1}}^{++}(\boldsymbol{q}) & 0 & 0 & u_{A_{1}}^{+-}(\boldsymbol{q})\\
0 & u_{B_{2}}^{++}(\boldsymbol{q}) & u_{B_{2}}^{+-}(\boldsymbol{q}) & 0\\
0 & u_{B_{2}}^{-+}(\boldsymbol{q}) & u_{B_{2}}^{--}(\boldsymbol{q}) & 0\\
u_{A_{1}}^{-+}(\boldsymbol{q}) & 0 & 0 & u_{A_{1}}^{--}(\boldsymbol{q})
\end{array}\right),
\end{align}
where $u_{\mathcal{A}}^{\xi\xi^{\prime}}(\boldsymbol{q})=\sum_{\boldsymbol{r}}u_{\mathcal{A}}(\boldsymbol{r})e^{i\boldsymbol{q}\cdot\boldsymbol{r}}e^{i(\mathbf{K}_{\xi}-\mathbf{K}_{\xi^{\prime}})\cdot\boldsymbol{r}}$.
By using 
\begin{align}
	\begin{split}
	&\langle u_{\mathcal{A}}^{\xi\xi^{\prime}} (\boldsymbol{q}) u_{\mathcal{A}^{\prime}}^{\mu\mu^{\prime}} (\boldsymbol{q}^{\prime}) \rangle  
	\\=& n_{\text{imp}}u^{2} \delta_{\mathcal{A}\mathcal{A}^{\prime}} \delta_{\mathbf{K}_{\xi} -\mathbf{K}_{\xi^{\prime}}, -(\mathbf{K}_{\mu}- \mathbf{K}_{\mu^{\prime}})} \delta_{\boldsymbol{q}, -\boldsymbol{q}^{\prime}},
	\end{split}
\end{align}
we can obtain the disorder averaging of the correlation function of disorder potential as
\begin{align}
\langle V_{\boldsymbol{q}}\otimes V_{\boldsymbol{q}^{\prime}}\rangle=\delta_{\boldsymbol{q},-\boldsymbol{q}^{\prime}}\sum_{\zeta,\kappa}\frac{n_{\text{imp}}u_{\zeta,\kappa}^{2}}{2}\varPi_{\zeta}\sigma_{\kappa}\otimes\varPi_{\zeta}\sigma_{\kappa},
\end{align}
with $u_{\zeta,\kappa}^{2}=u^{2}$ for $\zeta=\kappa=0$ or
$z$ and $u_{\zeta,\kappa}^{2}=u^{2}/2$ for $\zeta,\kappa=x,y$.
Terms $n_{\text{imp}}u_{x/y,x/y}^{2}$ take into account the inter-valley scattering. $n_{\text{imp}}u_{z,z}^{2}$
describes the different on-site energies for two layers. Term $n_{\text{imp}}u_{00}^{2}$ plays
the role of layer-symmetric disorder potential. The disordered averaged single particle
Green's function 
\begin{align}
G^{R/A}(\boldsymbol{p},\epsilon) & =\frac{E_{R/A}+\frac{1}{2m}[(p_{x}^{2}-p_{y}^{2})\sigma_{x}+2p_{x}p_{y}\sigma_{y}]}{E_{R/A}^{2}-p^{4}/(2m)^{2}},
\end{align}
with 
\begin{align}
	E_{R/A}=E-\Sigma^{R/A} ,
\end{align}  
where $\Sigma^{R/A}$ are the retarded and advanced self-energies from numerical calculations.
The velocity vertices renormalization by impurity scattering accounts for the ladder
diagrams shown in Fig.~\ref{fig:Feynman}(b) and can be obtained through the self-consistent equation, 
\begin{align}
\widetilde{\boldsymbol{v}}(\boldsymbol{p})=\boldsymbol{v}(\boldsymbol{p})+\sum_{s,l}n_{\text{imp}}u_{sl}^{2}\int\frac{d^{2}\boldsymbol{p}^{\prime}}{(2\pi)^{2}}G^{R}(\boldsymbol{p}^{\prime})\widetilde{\boldsymbol{v}}(\boldsymbol{p}^{\prime})G^{A}(\boldsymbol{p}^{\prime}).
\end{align}
It can be solved by assuming the renormalized vertex correction of the form $\widetilde{\boldsymbol{v}}(\boldsymbol{p})=\frac{1}{m^{*}}(p_{x}\sigma_{x}+p_{y}\sigma_{y},-p_{y}\sigma_{x}+p_{x}\sigma_{y})$
with $m^{*}$ the renormalized mass, and the second term in the right hand side vanishes
after the angular integration. Hence for the uncorrelated disorder, velocity vertices
are not renormalized.

The weak localization correction to the conductivity is associated with disorder-averaged
Cooperon function $C_{\mu\beta,\mu^{\prime}\beta^{\prime}}^{\xi\alpha,\xi^{\prime}\alpha^{\prime}}$
. The superscripts $\xi\alpha$ and $\xi^{\prime}\alpha^{\prime}$ indicate the valley
indices and the sublattice indices for the incoming and outgoing states in the retarded
branch, while the subscripts $\mu\beta$ and $\mu^{\prime}\beta^{\prime}$ are the
corresponding indices in the advanced branch. As illustrated diagrammatically in
Fig. \ref{fig:Feynman}(c), it can be evaluated from the Bethe-Salpeter (BS) equations
\begin{align}
	\begin{split}
& C_{\mu\mu^{\prime}, \beta\beta^{\prime}}^{\xi\xi^{\prime}, \alpha\alpha^{\prime}}(\boldsymbol{q})
\\=& n_{\text{imp}}u_{\zeta\kappa}^{2} \varPi_{\kappa}^{\xi\xi^{\prime}} \sigma_{\zeta}^{\alpha\alpha^{\prime}} \varPi_{\kappa}^{\mu\mu^{\prime}} \sigma_{\zeta}^{\beta\beta^{\prime}} \\&+\int\frac{d^{2}p}{(2\pi)^{2}} n_{\text{imp}}u_{\zeta\kappa}^{2} \varPi_{\kappa}^{\xi\lambda}\sigma_{\zeta}^{\alpha\gamma} \varPi_{\kappa}^{\mu\nu} \sigma_{\zeta}^{\beta\delta} 
\end{split}
\\& \times  G_{\lambda\lambda,\gamma\gamma^{\prime}}^{R}(\boldsymbol{p},\omega+\epsilon)G_{\nu\nu,\delta\delta^{\prime}}^{A}(\boldsymbol{q-p},\epsilon)C_{\nu\mu^{\prime},\delta^{\prime}\beta^{\prime}}^{\lambda\xi^{\prime},\gamma^{\prime}\alpha^{\prime}}(\boldsymbol{q}). \nonumber
\end{align}
The repeated indices obey the Einstein summation convention. It is convenient to
classify Cooperons as isospin (sublattice) and pseudospin (valley) singlets ($s,s^{\prime},l,l^{\prime}=0$)
and triplets ($s,s^{\prime},l,l^{\prime}=x,y,z$),
\begin{align}
C_{ll^{\prime}}^{ss^{\prime}}=\frac{1}{4}(\sigma_{y}\sigma_{s})_{\alpha\beta}(\varPi_{y}\varPi_{l})_{\xi\mu}C_{\mu\mu^{\prime},\beta\beta^{\prime}}^{\xi\xi^{\prime},\alpha\alpha^{\prime}}(\sigma_{s^{\prime}}\sigma_{y})_{\beta^{\prime}\alpha^{\prime}}(\varPi_{l^{\prime}}\varPi_{y})_{\mu^{\prime}\xi^{\prime}},
\end{align}
or inversely 
\begin{align}
	\begin{split}
C_{\mu\mu^{\prime},\beta\beta^{\prime}}^{\xi\xi^{\prime},\alpha\alpha^{\prime}}=& \frac{1}{4}C_{mm^{\prime}}^{rr^{\prime}}(\varPi_{m}\varPi_{y})_{\mu\xi}(\sigma_{r}\sigma_{y})_{\beta\alpha}
\\& \times (\varPi_{y}\varPi_{m^{\prime}})_{\xi^{\prime} \mu^{\prime}}(\sigma_{y}\sigma_{r^{\prime}})_{\alpha^{\prime}\beta^{\prime}}.
\end{split}
\end{align}
The impurity scattering vertex can be arranged in the same way 
\begin{align}
	\begin{split}
& n_{\text{imp}}u_{\zeta\kappa}^{2}\varPi_{\kappa}^{\xi\xi^{\prime}}\sigma_{\zeta}^{\alpha\alpha^{\prime}}\varPi_{\kappa}^{\mu\mu^{\prime}}\sigma_{\zeta}^{\beta\beta^{\prime}}\\
= & \frac{1}{4}n_{\text{imp}}u_{sl}^{2}(\sigma_{s}\sigma_{y})_{\beta\alpha}(\varPi_{l}\varPi_{y})_{\mu\xi}(\sigma_{y}\sigma_{s})_{\alpha^{\prime}\beta^{\prime}}(\varPi_{l}\varPi_{y})_{\xi^{\prime}\mu^{\prime}}
\end{split}
\end{align}
with 
\begin{align}
n_{\text{imp}}u_{sl}^{2}=\left(\begin{array}{ccccc}
& s=0 & x & y & z\\
l=0 & 2n_{\text{imp}}u^{2} & 0 & 0 & 0\\
x & 0 & n_{\text{imp}}u^{2} & n_{\text{imp}}u^{2} & 0\\
y & 0 & n_{\text{imp}}u^{2} & n_{\text{imp}}u^{2} & 0\\
z & 0 & 0 & 0 & 2n_{\text{imp}}u^{2}
\end{array}\right).
\end{align}
Then the BS equations for Cooperons in bilayer graphene read, 
\begin{align}
&C_{ll^{\prime}}^{ss^{\prime}}  =n_{\text{imp}}u_{sl}^{2}\delta_{ss^{\prime}}\delta_{ll^{\prime}}+\frac{1}{4}n_{\text{imp}}u_{sl}^{2}\int\frac{d^{2}p}{(2\pi)^{2}}\\
& \times\mathrm{Tr}\left\{ [G_{\boldsymbol{p},\omega+\epsilon}^{R}]^{\mathrm{T}}(\varPi_{y}\varPi_{l})(\sigma_{y}\sigma_{s})G_{\boldsymbol{q-p},\epsilon}^{A}(\varPi_{n}\varPi_{y})(\sigma_{t}\sigma_{y})\right\} C_{nl^{\prime}}^{ts^{\prime}}. \nonumber
\end{align}
Since the Green's function is diagonal in valley space, it leads to a series of coupled
equations for the Cooperon modes $C_{ll}^{ss^{\prime}}\equiv C_{l}^{ss^{\prime}}$, and
\begin{align}
C_{l}^{ss^{\prime}}  =& n_{\text{imp}}u_{sl}^{2}\delta_{ss^{\prime}}+n_{\text{imp}}u_{sl}^{2}\int\frac{d^{2}p}{(2\pi)^{2}}\label{eq:cooperon_valley}\\
& \times\frac{1}{2}\mathrm{Tr}\left\{ [G_{\boldsymbol{p},\epsilon}^{R}]^{\mathrm{T}}(\sigma_{y}\sigma_{s})G_{\boldsymbol{q-p},\epsilon}^{A}(\sigma_{t}\sigma_{y})\right\} C_{l}^{ts^{\prime}},\nonumber 
\end{align}
where the trace is only evaluated in sublattice space. The BS equations can be solved
by using the gradient expansion of $G^{A}$ in the small wavevector $\boldsymbol{q}$
\begin{align}
G_{\boldsymbol{q-p},\epsilon}^{A}  \simeq &\ G_{-\boldsymbol{p},\epsilon}^{A} +G_{-\boldsymbol{p},\epsilon}^{A}(\boldsymbol{q}\cdot\boldsymbol{v}_{-\boldsymbol{p}})G_{-\boldsymbol{p},\epsilon}^{A}\nonumber \\
& +G_{-\boldsymbol{p},\epsilon}^{A}(\boldsymbol{q}\cdot\boldsymbol{v}_{-\boldsymbol{p}})G_{-\boldsymbol{p},\epsilon}^{A}(\boldsymbol{q}\cdot\boldsymbol{v}_{-\boldsymbol{p}})G_{-\boldsymbol{p},\epsilon}^{A}\label{eq:gradient_expansion_GA}\\
& +\frac{1}{2}G_{-\boldsymbol{p},\epsilon}^{A}(q_{i}v_{-\boldsymbol{p}}^{ij}q_{j})G_{-\boldsymbol{p},\epsilon}^{A},\nonumber 
\end{align}
where $v_{\boldsymbol{p}}^{ij}=\frac{\partial^{2}H}{\partial p_{i}\partial p_{j}}$.
The zero order in $\boldsymbol{q}$ determines the relaxation gap for each Cooperon
channel
\begin{align}
& \int\frac{d^{2}p}{(2\pi)^{2}}\frac{1}{2}\mathrm{Tr}\left\{ [G_{\boldsymbol{p},E}^{R}]^{\mathrm{T}}(\sigma_{y}\sigma_{s})G_{\boldsymbol{-p},E}^{A}(\sigma_{s^{\prime}}\sigma_{y})\right\} \nonumber \\
 &= \int_{0}^{\infty}\frac{pdp}{2\pi}\frac{1}{[E_{R}^{2}-(\frac{p^{2}}{2m})^{2}][E_{A}^{2}-(\frac{p^{2}}{2m})^{2}]} \\
& \times\left(\begin{array}{ccccc}
	& s=0 & x & y & z\\
	s^{\prime}=0 & E_{R}E_{A}-\frac{p^{4}}{4m^{2}} & 0 & 0 & 0\\
	x & 0 & E_{A}E_{R} & 0 & 0\\
	y & 0 & 0 & E_{A}E_{R} & 0\\
	z & 0 & 0 & 0 & E_{A}E_{R}+\frac{p^{4}}{4m^{2}}
\end{array}\right).\nonumber
\label{eq:cooperon_gap}
\end{align}
Due to the quadratic spectrum of Hamiltonian (\ref{eq:Hamiltonian}) and $\boldsymbol{v}_{\boldsymbol{p}}$ with linear momentum, the linear order in $\boldsymbol{q}$ vanishes.
The couplings between different channels from $\boldsymbol{q}^{2}$ terms give a
higher power of $q$ contribution and can be neglected. As a consequence, the singlet-triplet
channel in sublattice space is also conserved after multi-scattering $C_{l}^{ss^{\prime}}=\delta_{ss^{\prime}}C_{l}^{s}$.
From Eq. (\ref{eq:cooperon_valley}), we have
\begin{equation}
C_{l}^{s}=\frac{n_{\text{imp}}u_{sl}^{2}}{1-\frac{n_{\text{imp}}u_{sl}^{2}}{2}\int\frac{d^{2}p}{(2\pi)^{2}}\mathrm{Tr}\left\{ [G_{\boldsymbol{p},\epsilon}^{R}]^{\mathrm{T}}(\sigma_{y}\sigma_{s})G_{\boldsymbol{q-p},\epsilon}^{A}(\sigma_{s}\sigma_{y})\right\} }.\label{eq:valley_sublattice_cooperon}
\end{equation}
Within the self-consistent Born approximation, the self-energy can be obtained by
solving the following self-consistent equation,
\begin{align}
	\begin{split}
\varSigma_{\boldsymbol{p},E}^{R/A} & =\int\frac{d^{2}p^{\prime}}{(2\pi)^{2}}\langle V_{\boldsymbol{p}-\boldsymbol{p}^{\prime}}G_{\boldsymbol{p}^{\prime},\epsilon}^{R/A}V_{\boldsymbol{p}^{\prime}-\boldsymbol{p}}\rangle\\
& =2n_{\text{imp}}u^{2}\int\frac{d^{2}p^{\prime}}{(2\pi)^{2}}\frac{E_{R/A}}{E_{R/A}^{2}-p^{4}/(2m)^{2}},
\end{split}
\end{align}
which yields the Ward identity
\begin{align}
	\begin{split}
1 & =\frac{\varSigma^{R}(\mathbf{p},\epsilon)-\varSigma^{A}(\mathbf{p},E)}{(E_{A}-E_{R})}\\
& =2n_{\text{imp}}u^{2}\int\frac{d^{2}p^{\prime}}{(2\pi)^{2}}\frac{E_{A}E_{R}+(\frac{p^{2}}{2m})^{2}}{[E_{R}^{2}-(\frac{p^{2}}{2m})^{2}][E_{A}^{2}-(\frac{p^{2}}{2m})^{2}]}.
\end{split}
\end{align}
The analysis shows that the only Cooperon channel that remains gapless is the sublattice-triplet
and valley-triplet Cooperon $C_{z}^{z}$, which belongs to the intervally channel
category. The intravalley Cooperon channels $C_{l}^{x,y}$ are strongly suppressed
by the intervalley scattering from the atomically sharp scatters. Here we want to
emphasize the exact cancellation of the zero order term in $\boldsymbol{q}$ in the
denominator of $C_{z}^{z}$ is ensured by Ward identity, regardless of the explicit
form of the self-energy $\Sigma^{R/A}$. After substituting Eq. (\ref{eq:gradient_expansion_GA})
into Eq. (\ref{eq:valley_sublattice_cooperon}) and performing the integral, we have
\begin{align}
C_{z}^{z}=\frac{2n_{\text{imp}}u^{2}}{l_{e}^{2}q^{2}}
\end{align}
with the square of mean free path defined as
\begin{align}
l_{e}^{2}=-\frac{n_{\text{imp}}u^{2}}{\pi}\frac{\left(E_{A}E_{R}\left(\ln\left(-E_{A}^{2}\right)-\ln\left(-E_{R}^{2}\right)\right)-E_{A}^{2}+E_{R}^{2}\right)}{\left(E_{A}-E_{R}\right){}^{3}\left(E_{A}+E_{R}\right)}.
\end{align}
The leading quantum correction to the conductivity can be computed by an index contraction
of the external legs of the Cooperon with the Hikami boxes,
\begin{align}
\sigma_{\text{qi}}= & \frac{1}{2\pi}\frac{e^{2}}{\hbar}\frac{1}{V^{2}}\sum_{\boldsymbol{k},\boldsymbol{q}}(G_{\boldsymbol{k}}^{A}v_{\boldsymbol{k}}^{x}G_{\boldsymbol{k}}^{R})_{\mu^{\prime}\beta^{\prime},\xi\alpha}\\
& \times C_{\mu\mu^{\prime},\beta\beta^{\prime}}^{\xi\xi^{\prime},\alpha\alpha^{\prime}}(\boldsymbol{k},-\boldsymbol{k},\boldsymbol{q})(G_{\boldsymbol{q}-\boldsymbol{k}}^{R}v_{\boldsymbol{q}-\boldsymbol{k}}^{x}G_{\boldsymbol{q}-\boldsymbol{k}}^{A})_{\xi^{\prime}\alpha^{\prime},\mu\beta}.
\nonumber
\end{align}
The bare Hikami box depicted as the second diagram in Fig. \ref{fig:Feynman}(d)
is enough. The two other dressed diagrams as shown in Fig. \ref{fig:Feynman}(e)
which must be included in single layer graphene vanishes for bilayer graphene since
$\boldsymbol{v}_{\boldsymbol{p}}$ is linear in momentum. After being transformed
into singlet and triplets channels, 
\begin{align}
\sigma_{\text{qi}} & \simeq\frac{1}{2\pi}\frac{e^{2}}{\hbar}\mathcal{H}_{rr^{\prime}}^{mm^{\prime}}\frac{1}{V}\sum_{\boldsymbol{q}}C_{mm^{\prime}}^{rr^{\prime}}(\boldsymbol{q})
\end{align}
with $\mathcal{H}_{rr^{\prime}}^{mm^{\prime}}$ as the Hikami box for each channel
\begin{align}
&\mathcal{H}_{rr^{\prime}}^{mm^{\prime}}  =\delta_{mm^{\prime}}\delta_{rr^{\prime}}\frac{1}{2}\mathrm{Tr}[\varPi_{y}\varPi_{m^{\prime}}\varPi_{y}^{T}\varPi_{m}^{T}] \frac{1}{V}
\\& \times
\sum_{\boldsymbol{k}}\frac{1}{2}\mathrm{Tr}\left[(\sigma_{y}\sigma_{r})G_{\boldsymbol{k}}^{A}v_{\boldsymbol{k}}^{x}G_{\boldsymbol{k}}^{R}(\sigma_{r}\sigma_{y})^{\mathrm{T}}(G_{-\boldsymbol{k}}^{R}v_{-\boldsymbol{k}}^{x}G_{-\boldsymbol{k}}^{A})^{\mathrm{T}}\right].
\nonumber
\end{align}
The Hikami box for $C_{z}^{z}$ can be evaluated as 
\begin{align}
\mathcal{H}_{z}^{z} & =-\frac{1}{\pi}\frac{\left(E_{A}E_{R}\left(\ln\left(-E_{A}^{2}\right)-\ln\left(-E_{R}^{2}\right)\right)-E_{A}^{2}+E_{R}^{2}\right)}{\left(E_{A}-E_{R}\right)^{3}\left(E_{A}+E_{R}\right)}.\label{eq:Hikami_box}
\end{align}
The lower $q$ cutoff for a system of length $L$ is $\sim1/\min\{L,L_{\phi}\}$,
where $L_{\phi}$ is the coherence length of the system, and the upper cutoff is $\sim1/\ell_{e}$.
The quantum interference conductivity correction is negative (weak localization)
and can be evaluated as 
\begin{align}
\sigma_{\text{qi}}\simeq-\frac{e^{2}}{\pi h}\ln\frac{\min\{L,L_{\phi}\}}{\ell_{e}}.
\end{align}

	\underline{}
	
	%

\end{document}